\documentclass[12pt,dvips]{article}

\usepackage{amsmath,amssymb,exscale}
\usepackage{array,multicol}
\usepackage{afterpage,float,flafter}
\usepackage{epsfig,rotating,pifont,fancybox}
\usepackage{cite}
\makeatletter
\@addtoreset{equation}{section}
\makeatother

\title{
\vspace{1cm} 
\Large\textbf{
RS1, Custodial Isospin and Precision Tests
}
\vspace*{.5cm}
\author{\large \textbf{
Kaustubh Agashe\footnote{email: kagashe@pha.jhu.edu} , 
Antonio Delgado\footnote{email: adelgado@pha.jhu.edu},}\\
\textbf{Michael J. May\footnote{email: mjmay@pha.jhu.edu}
\mbox{  }and Raman Sundrum\footnote{email: sundrum@pha.jhu.edu}}\\
\emph{
Department of Physics and Astronomy} \\ 
\emph{Johns Hopkins University} \\ 
\emph{3400 North Charles St}. \\ 
\emph{Baltimore, MD 21218-2686}}}

\date{}
\def\be{\begin{equation}}
\def\ee{\end{equation}}

\begin{document}
\maketitle
\thispagestyle{empty}
\vspace*{.5cm}
  
\begin{abstract} 

We study precision electroweak constraints within a RS1 model
with gauge fields and fermions in the bulk. The electroweak gauge symmetry
is enhanced to $SU(2)_L \times SU(2)_R \times U(1)_{B-L}$,
thereby providing a custodial isospin symmetry 
sufficient to suppress excessive contributions to the $T$ 
parameter. We then construct complete models, complying with
all electroweak constraints,
for solving the hierarchy problem, without supersymmetry or large
hierarchies in the fundamental couplings. Using the
AdS/CFT correspondence our models can be interpreted as dual to a strongly
coupled conformal Higgs sector with \emph{global} custodial 
isospin symmetry, gauge 
and fermionic matter
being fundamental fields external to the CFT. This scenario has interesting 
collider signals, distinct from other RS models in the literature. 

\end{abstract} 
  
\newpage 
\renewcommand{\thepage}{\arabic{page}} 
\setcounter{page}{1} 

\section {Introduction}

There is a puzzle at the heart of particle physics which has become ever 
sharper in the last two decades of experimental and theoretical research.
The minimal Standard Model (SM) is thusfar 
in superb agreement with experiment, not 
just in terms of the central functions for which it was designed, but 
remarkably,  in 
every accidental detail following from its minimality, such as 
suppressed flavor-changing neutral currents (FCNC's), proton stability, 
and a host of precision electroweak effects. Yet undeniably, the SM 
effective field theory suffers from the hierarchy problem and forces us 
to look beyond. Any 
approach for solving the hierarchy problem involves extending  the 
 SM at the weak scale 
and, in one way or another,  
threatens the economical and detailed agreement with experiment. 
Given this fundamental tension it is of considerable importance to 
identify within the different approaches to the hierarchy problem, robust
effective field theory mechanisms which protect the key features of 
particle  phenomenology,  as well as the future experimental implications of
these mechanisms.

The Randall-Sundrum I model (RS1) \cite{rs1}  \cite{gw} 
presented an exciting
approach to the 
hierarchy problem based on geometrical hierarchies arising from warped 
higher-dimensional spacetime. However, most of the finer but 
interesting phenomenological issues 
are sensitive to the UV completion of the 
original RS1 effective field theory. The AdS/CFT correspondence \cite{adscft}
offers a great deal of insight into the RS1 proposal \cite{rscft}. Via this 
correspondence, RS1 is dual to a 
purely 4D theory of particle physics and gravity, albeit one involving a 
strongly-coupled sector which is conformally invariant between the Planck 
and weak scales. The RS1 Kaluza-Klein excitations as well as fields localized 
on the ``IR'' brane are interpreted as TeV-scale composites of the strong 
sector. 
Fundamental fields coupled to strong CFT operators appear together as 
bulk RS fields. 
In the original RS1 model, all SM fields are localized on the 
IR brane, and therefore the model is dual to TeV-scale compositeness of the 
entire SM. The details of this compositeness determine the fate of the 
various
phenomenological questions, but they are dual to details of the unknown 
UV completion of the RS1 effective theory. 

There is another direction one can take. On the 4D side, to
solve the hierarchy problem it
is sufficient for just 
the Higgs to be a TeV-scale composite of a strongly interacting sector 
\cite{comphiggs}, the masses of 
higher-spin fundamental fields being protected by chiral or gauge symmetries.
 Of course for gauge boson and fermion masses to arise at 
the weak scale, the fundamental fields must couple to the Higgs sector. 
There is a simple way of studying this in the dual RS setting by 
continuing to take the Higgs to be localized on the IR brane, but taking 
gauge bosons and fermions to propagate in the higher dimensional bulk.
The great advantage of doing this is that the key phenomenological issues 
become IR-dominated, and therefore addressable, in weakly-coupled 
RS effective field theory, rather than being sensitive to its UV completion.
We find this approach very exciting and promising.

Let us briefly review the history of such studies. With bulk gauge fields, 
the calculability of weak scale effects at first seemed a liability, with 
large harmful effects for 
compositeness \cite{ewbad} \cite{shafi} \cite{csaki} \cite{hewett2} \cite{burdman}
and
precision electroweak observables 
\cite{shafi} \cite{csaki} \cite{hewett2} \cite{burdman}.
Reference \cite{csaki} presented their results in terms of the Peskin-Takeuchi
$S$ and $T$ parameters \cite{pt}, which facilitated a global fit to the data.
It was later 
realized that placing fermion fields in the 
bulk allowed one to greatly soften some of these effects \cite{gp} 
\cite{shafi} \cite{hewett1} \cite{carena2}. There were further 
dividends in that bulk fermion masses provided a simple 
attractive mechanism for 
generating Yukawa structure without fundamental hierarchies in the RS1 action
\cite{neubert} \cite{gp} \cite{huber}.
Furthermore the same mechanism automatically protects the theory from 
excessive FCNC's \cite{gp} \cite{huber}. 
The issue of gauge coupling running and unification was 
discussed in Refs. \cite{running} \cite{lisa} \cite{choi} \cite{us}, 
with complete models with unification constructed in Ref. \cite{us}.
In particular a mechanism for protecting baryon stability was given in 
reference \cite{us}, adapting some key features of the mechanism of reference \cite{gns}.

The last major phenomenological obstacle remaining in this program of research
has been the problem of excessive contributions 
\cite{shafi} \cite{csaki} \cite{hewett2} 
to the Peskin-Takeuchi $T$ parameter \cite{pt}. 
The usual model-building rule for protecting 
this parameter is to ensure that the Higgs sector, when considered in 
isolation from gauge and fermion fields, should have a custodial isospin 
symmetry after electroweak symmetry breaking, under which the $W$'s form 
a triplet. However, the various RS1 models studied already appear to comply
with this rule, since they make use of the minimal Higgs on the IR brane. 
However, the problem can be identified when one views these models from the 
dual 4D perspective. Bulk RS gauge fields are dual to both 
fundamental 4D gauge fields and to the 
CFT operators to which they might couple, 
namely conserved {\em global} symmetry 
currents of the CFT. This CFT represents the entire 
Higgs sector on the 4D side, of which the minimal Higgs is a 
light composite. The dual of
the {\em entire} CFT Higgs sector enjoying a {\em global} 
custodial isospin symmetry is therefore to have a custodial isospin 
{\it gauge} symmetry in the RS {\em bulk}. 
Earlier RS models focused on {\em just} the SM gauge symmetry in the bulk. From 
the dual point of view, their difficulties with the $T$ parameter
trace to the
absence of custodial isospin symmetry in the CFT Higgs sector.

In this paper we study just such a 
bulk custodial isospin scenario and show that this extra gauge 
symmetry protects the $T$ parameter
adequately. We are then able to construct fully realistic models satisfying 
all precision electroweak and other constraints. This is significant 
because we accomplish this in a non-supersymmetric approach to the 
hierarchy problem and without 
invoking any large fundamental hierarchies.
In a composite Higgs model, the scale of compositeness can be made a free 
parameter. It can be raised at the cost of fine-tuning in the sense of the 
hierarchy problem, or lowered at the cost of strong interactions becoming 
more phenomenologically dangerous. The same is true in our RS model, where 
the scale of Kaluza-Klein resonances is dual to the compositeness scale. 
Our fit to the body of precision test data requires such resonances to be 
above about $3$ TeV. 

An important consideration in this fit arises from
the third generation quarks, especially the
tension between the need to generate a large top quark mass while 
suppressing large corrections to bottom quark couplings to the $Z$.
While some of the collider signals of our model are familiar expectations 
of strong interactions above the weak scale, some are more distinctively 
 linked  to the third generation constraints. 

Our study illustrates the utility of RS effective field theory as a 
weakly coupled approach to a traditionally strongly-coupled and difficult 
subject: the 
possibility that the hierarchy problem is solved by non-supersymmetric 
physics above the weak scale. It allows us to calculate the signs and sizes
of the leading effects on interesting observables in terms of model inputs, 
rather than just rough estimates.
RS calculability is bought at a price. 
The dual strongly coupled theory must have special features \cite{adscft} 
\cite{rscft}: it must 
be approximately conformally invariant over the Planck-weak hierarchy, 
have  a large-$N$ type expansion, and have a large gap in the 
spectrum of CFT scaling dimensions, with only a finite number 
being close to marginal  (which
automatically includes any symmetry CFT currents). Nevertheless, we have 
found that these special features do not pose any extra phenomenological 
liability, and are indeed an asset. Further, we expect 
many of our conclusions to survive even if some of the above theoretical 
control parameters are relaxed in Nature. 

This paper is organized as follows. Section 2 describes the set-up of our 
model. Section 3 is a brief discussion of electroweak precision variables and 
the subtleties peculiar to our model. Sections 4 and 5 derive the tree level  
contributions to the Peskin-Takeuchi $S$ and $T$ parameters \cite{pt}. Section 6  
derives the top loop contribution to $T$ in our model, bulk custodial 
isospin ensuring UV finiteness. Section 7 shows how our model can naturally 
fit the electroweak data. Section 8 describes the central new collider 
signals of our model. Section 9 describes the 4D dual CFT interpretation of 
our model and results. Section 10 provides further discussion and the 
outlook for future progress in this arena. Many of the more technical 
details have been relegated to appendices.  

\section{The Model}
\label{model}
\subsection{Overview}

We are going to study a model with
$SU(3)_c \times SU(2)_L\times SU(2)_R\times U(1)_{B-L}$ gauge symmetry in the bulk
of a warped extra dimension. In order to recover the usual 
$SU(2)_L\times U(1)_Y$ we will break $SU(2)_R$ with orbifold boundary conditions
on the Planck brane to $U(1)_R$, 
keeping the IR brane $SU(2)_R$ symmetric. 
Then we will break $U(1)_R\times U(1)_{ B_ L }
\rightarrow U(1)_Y$ spontaneously on the Planck brane.
In one of the scenarios we consider, we will further break
$SU(2)_R$ in the bulk by a small amount.

The metric of RS1 can be written as:
\begin{equation}
( d s) ^2  = \frac{1}{ ( kz )^2 } \big[ \eta_{ \mu \nu }
d x ^{ \mu } d x^{ \nu } - ( d z ) ^2 \big]. 
\end{equation}
Here,
\be
\left( z_h \equiv \frac{1}{k} \right) 
\leq  z \leq \left( z_v \equiv \frac{ e^{ k \pi r_c }}{k} \right),
\ee
where $k$ represents the $AdS_5$ curvature, $z_v \sim \mbox{TeV}^{-1}$, and $\eta_{\mu\nu}$
is the 4D Minkowski metric.  We take
\be
k \pi r_c \sim \log \left( M_{Pl} / TeV 
\right) \sim 30
\ee
to solve the hierarchy problem. 

In that background the lagrangian for our
model reads:
\begin{equation}
S=\int d^4x dz \sqrt{G}
(\mathcal{L}_{gauge}+
\mathcal{L}_{fermion} + 
\mathcal{L}_{\mathrm{UV}}\delta(z-z_h)+\mathcal{L}_{\mathrm{IR}}
\delta(z-z_v)).
\end{equation}
\noindent
$\mathcal{L}_{gauge} + \mathcal{L}_{fermion}$ is the bulk lagrangian.
We focus on $\mathcal{L}_{gauge}$ first,  discussing
 $\mathcal{L}_{fermion}$ in section 2.4.
\begin{eqnarray}
\mathcal{L}_{gauge}&=&\sqrt{G}\big(-\frac{1}{4} Tr W_{MN}
W^{MN}
-\frac{1}{4} Tr \widetilde{W}_{MN}\widetilde{W}^{MN} \nonumber\\
&& -\frac{1}{4} Tr \widetilde{B}_{MN}\widetilde{B}^{MN} 
-\frac{1}{4} Tr F_{MN}
F^{MN} 
+ | D_M \Sigma |^2 - V (\Sigma) \big),
\label{bulk}
\end{eqnarray}
where the indices are contracted with the bulk metric $G_{MN}$, and $W^{MN}$
is field strength for the $SU(2)_L$ gauge group, $\widetilde{W}_{MN}$ for $SU(2)_R$,
$\widetilde{B}_{MN}$ for $U(1)_{B-L}$ and $F_{MN}$ is for the gluon.  
$\Sigma$ is a triplet of $SU(2)_R$ whose sole purpose is to spontaneously
break $SU(2)_R$ to $U(1)_R$ at a mass scale below $k$. Therefore,
henceforth, we will simply work with the gauge theory with a mass term for
$\tilde{W}^{ \pm }$:
\begin{eqnarray}
\mathcal{L}_{gauge}&=&\sqrt{G}\big(-\frac{1}{4} Tr W_{MN}
W^{MN}
-\frac{1}{4} Tr \widetilde{W}_{MN}\widetilde{W}^{MN} \nonumber\\
&& -\frac{1}{4} Tr \widetilde{B}_{MN}\widetilde{B}^{MN}  
-\frac{1}{4} Tr F_{MN}
F^{MN} + \tilde{M}^2 | \tilde{W}^{ \pm } |^2 \big)
\end{eqnarray}
%
In fact, it will be interesting to consider
$2$ separate cases, {\em Scenario I}, 
where $\tilde{M} /k$ is small, but non-zero and {\em Scenario II},
where  $\tilde{M} /k = 0$, i.e., $SU(2)_R$ is unbroken in the bulk.

$\mathcal{L}_{\mathrm{UV}}$ includes the necessary fields to spontaneously break
$U(1)_R \times
U(1)_{B-L}$ to $U(1)_Y$
and $\mathcal{L}_{\mathrm{IR}}$ contains the SM Higgs field,
now a {\em bi}doublet of $SU(2)_L \times SU(2)_R$:
\begin{equation}
\mathcal{L}_{\mathrm{IR}}= \mathcal{L}_{Higgs} + \mathcal{L}_{Yukawa},
\end{equation}
where
${\cal L}_{Yukawa}$ will generate Yukawa couplings for
fermions which will be discussed in section \ref{fermionmodel} and
\begin{equation}
\mathcal{L}_{Higgs}=
\sqrt{-g_{\mathrm{IR}}} \left( D_\mu H \big[ D^\mu H \big] ^{ \dagger } -V( H )\right).
\label{LHiggs}
\end{equation}
$g_{IR}$ is the \emph{induced} flat space metric in the IR brane.  
After the usual field redefinition of $H$ \cite{rs1},
Eq.~(\ref{LHiggs}) takes its canonical form:
\begin{equation}
\mathcal{L}_{Higgs}= D_\mu H \big[ D^\mu H \big] ^{ \dagger } -V( H )
\label{LHiggscan}
\end{equation}
%
with $\langle H \rangle = 
\left( 
\begin{array}{c}
0 \\
v / \sqrt{2}
\end{array}
\right)$, $v \approx 250$GeV, and 
the ratio of the Higgs vev to the warped down curvature scale is taken as
\be
v z_v \sim \frac{1}{5}. 
\ee

We assume that brane-localized (kinetic) terms for bulk
fields are of order loop processes involving bulk couplings 
and are therefore neglected in our analysis (however, see references 
\cite{hewett3, carena1, carena2} for effects of larger
brane-localized kinetic terms for gauge fields and reference \cite{santiago}
for effects of brane-localized kinetic terms for fermions.).

\subsection{$SU(2)_R \rightarrow U(1)_R$ by orbifold boundary condition}

In addition to bulk breaking of $SU(2)_R$, we further break $SU(2)_R$ to $U(1)_R$
by orbifold boundary condition.
Therefore,
we assign the following 
boundary conditions to the $\mu$-components of the gauge fields 
under an  $S^1 / \mathbb{Z}_2\times\mathbb{Z}^\prime_2$ orbifold 
\cite{orbifoldnew, bhn, jmr}. 
\[
\begin{array}[t]{ccc}
 & \textrm{UV} & \textrm{IR}  \\
\widetilde{W}_{\mu}^{1,2}  & - & + \\
\hbox{other gauge fields} & + & + 
\end{array}
\]

\subsection{$U(1)_R \times U(1)_{B-L} \rightarrow U(1)_Y$ on UV brane}
\label{3rb-lmix}

The breaking of 
$U(1)_R \times U(1)_{B-L} \rightarrow U(1)_Y$ occurs via a vev on the UV brane. 
There are two linear combinations of $\widetilde{W}_{\mu}^3$ and $\widetilde{B}_\mu$,
\be
Z^{\prime}_\mu \equiv \frac{\tilde{g}_5 \widetilde{W}_{\mu}^3-\tilde{g}_5^\prime 
\widetilde{B}_\mu}{\sqrt{\tilde{g}_5^2+\tilde{g}_5^{\prime 2}}}, \qquad \mathrm{and} \qquad  
B_\mu \equiv  \frac{\tilde{g}_5^\prime \widetilde{W}_{\mu}^3 + \tilde{g}_5 
\widetilde{B}_\mu}{\sqrt{\tilde{g}_5^2+\tilde{g}_5^{\prime2}}}, 
\ee
where
\be
D_M = 
\partial_M-i (  g_5 W^a_{M}
\tau_{a \, L} + \tilde{g}_5 \widetilde{W}^a_M \tau_{a \, R} + \tilde{g}_5^\prime 
\widetilde{B}_M \widetilde{Y} ), 
\ee
is the electroweak covariant derivative
with $\widetilde{Y}= \frac{B-L}{2}$.
$B_\mu$ is the hypercharge gauge boson.  It is $(+,+)$. 
We couple $Z^{\prime}_\mu$ to a Planckian vev on the UV brane 
which mimics $(-, +)$ boundary condition to 
a good approximation.

In terms of $Z^{\prime}$ and $B$, the five dimensional 
electroweak
covariant derivative is now 
\begin{eqnarray}
\label{bzD}
D_M & = & \partial_M -i \big( g_{5} W_{M}^a \tau_{a \, L} + 
\tilde{g}_5 \widetilde{W}^{1,2}_{M} \tau^{1,2}_{R} + 
g_{5 Z^\prime} Z^{\prime}_M Q_{Z^\prime} + g_5^\prime B_M ( \tau^3_R + \widetilde{Y}) \big).
\end{eqnarray}
We have defined the hypercharge coupling,
\be
g^{ \prime }_5= \frac{ \tilde{g}^{ \prime }_5 \; 
\tilde{g}_5 }{ \sqrt{ \tilde{g}_5^2 + \tilde{g}_5^{\prime \; 2}}},
\ee
the $Z^\prime$ charge
\be
Q_{Z^\prime}= \tau^3_R-\sin^2 \theta^\prime \; Y,
\label{QZprime}
\ee
the $Z^\prime$ coupling
\be
g_{ Z^\prime \; 5 }= \sqrt{\tilde{g}_5^2+\tilde{g}_5^{\prime \; 2} },
\ee
and the $\widetilde{B}$ -- $\widetilde{W}^3$ mixing angle
\be
\sin \theta^\prime= \frac{ \tilde{g}_5^{ \prime } }{ g_{ Z^{ \prime } \; 5 } }.
\ee


\subsection{Fermions}
\label{fermionmodel}
Since we have an enhanced bulk gauge symmetry,
namely $SU(2)_R$, we have to promote the usual right handed fermionic
fields
to doublets of this symmetry. 
Moreover, since we are breaking that 
symmetry through the UV orbifold, one component of $SU(2)_R$ doublet
must be even and therefore has a zero-mode
while the other component must be odd and therefore does not have a zero-mode.
Therefore, 
we are forced to double the number
of right handed doublets in such a way that from one of them
the upper component, for example up type quark, is even
whereas from the other the lower component, down type, is even 
-- this is
similar to obtaining quarks and lepton zero-modes
from different $SU(5)$ bulk multiplets in orbifolded GUT scenarios 
\cite{bhn, jmr}. 
This doubling
of right handed particles is only needed in the quark sector, since
in the lepton sector we only need the right handed charge leptons\footnote{
Although in any embedding of this theory in a GUT, the minimal group
would be $SO(10)$, thus needing a $\nu_R$ and another doublet.} to be 
massless after we compactify.
Explicitly,
we have three types of doublets under
$SU(2)_R$ per generation in such a way that\footnote{Only one 
chirality will be discussed since the
other chirality is projected out by  $\mathbb{Z}_2$ symmetry.}: 
\begin{eqnarray}
Q_{R \; 1} & = & u_R + d^{ \prime }_R \nonumber \\
Q_{R \; 2} & = & u^{ \prime }_R + d_R \nonumber \\
L_R & = & e_R + \nu_R^{ \prime }
\end{eqnarray}
where the {\em un}primed particles are the ones to have zero modes, i.e. 
to be $(+,+)$. The extra 
fields needed to complete all representations
are $(-,+)$ 
(since breaking of $SU(2)_R$ is on the Planck brane).

The general bulk lagrangian for fermions is:
\begin{equation}
\mathcal{L}_{fermion}=\sqrt{G}(i\bar{\Psi}\Gamma^M D_M \Psi-\epsilon(y) c_{\Psi}
\bar{\Psi}\Psi)
\label{fermions}
\end{equation}
where $\epsilon(y)$ is the sign function. Even though it will seem
that we are adding a mass term, $c_{\Psi}$ is compatible with a massless zero
mode. This parameter controls the localization of the zero mode, for $c>1/2$
($c<1/2$) the wavefunction near the Planck (IR brane) \cite{neubert, gp}. 

The Yukawa
couplings to Higgs 
(prior to field redefinition of Eq.~(\ref{LHiggs}) $\rightarrow$ Eq.~(\ref{LHiggscan}))
are necessarily localized on the IR brane:
\begin{eqnarray}
\mathcal{L}_{Yukawa} & = & \sqrt{ - g_{IR} } 
H \left( \lambda_{ u \; 5} Q_L Q_{R \; 1} + 
\lambda_{ d \; 5} Q_L Q_{R \; 2}
+ \lambda_{ e \; 5} L_L L_R \right)
\end{eqnarray}
Note that because $u_R$ and $d_R$ zero-modes arise from
{\em different} $SU(2)_R$ doublets, we are able to give them 
separate Yukawa couplings
without violating $SU(2)_R$ on the IR brane.

So far, we have detailed the model, except for choice of $c$'s.
The $c$ parameters
give a simple mechanism for obtaining
hierarchical $4D$ 
Yukawa couplings with{\em out} hierarchies
in $5D$ Yukawa couplings. In short, light fermions
are localized near {\em Planck} brane ($c > 1/2$) so that their 
$4D$ Yukawa couplings are
small
due to the small overlap with Higgs on {\em TeV} brane. 
Left-handed top and bottom quarks are close to $c = 1/2$ 
(but $< 1/2$)\footnote{As 
we will show $c_L \sim 1/2$ is necessary to
be consistent with $Z \rightarrow \bar{b}_L b_L$
for KK masses $\sim$ few TeV.}, whereas {\em right}-handed top quark is localized 
near TeV brane to get $O(1)$ top
Yukawa. 
With this set-up, FCNC's 
from exchange of 
both gauge KK modes and ``string states'' (parameterized
by flavor-violating local operators in our effective field theory) are also 
suppressed.
See references \cite{gp, huber} for details.

\section{Electroweak precision observables}
\label{formalism}

We begin with formalism for electroweak fit in the presence of new physics.
It is convenient to discuss this in the framework of
$4D$ effective Lagrangian at the weak scale
for SM with all the heavy non-standard physics integrated out \cite{bs}
(as pioneered in
references \cite{georgi}, but here retaining the Higgs field).
This framework was used earlier in the RS model studied in
reference \cite{csaki}.
The dimension-$6$ operators, obtained by integrating out
heavy particles, which are important for the electroweak fit
are:
\begin{eqnarray}
{\cal L}_{ gauge-kinetic } & = & \frac{ g g^{ \prime } s }{ 16 \pi^2 v^2 }
H^{\dagger} \tau^a H B^{ \mu \nu}
W_{ a \; \mu \nu }
\label{kinetic}
\end{eqnarray}
\begin{eqnarray}
{\cal L}_{ gauge-mass } & = & 
\frac{-t}{ 16 \pi^2 v^2 } \big[ \left( D^{\mu} H \right)^{\dagger} H \big]
\left( H^{\dagger} D_{\mu} H \right)
\label{mass}
\end{eqnarray}
\begin{eqnarray}
{\cal L}_{fermion} & = & \frac{ - i x }{ 16 \pi^2 v^2 } 
\bar{\psi} \gamma^{\mu} \tau^a \psi
\left( D_{\mu} H \right)^{\dagger} \tau_a H + 
\frac{ - i y }{ 16 \pi^2 v^2 } \bar{\psi} \gamma^{\mu}
\psi \left( D_{\mu} H \right)^{\dagger} H + \frac{V}{ 16 \pi^2 v^2 }
\bar{\psi} \psi \bar{\psi} \psi + h.c., \nonumber \\
\label{fermionoperator}
\end{eqnarray}
where $x$, $y$ and $V$, in general, vary with the fermion.

Usually, the gauge-kinetic 
higher-dimensional operator in Eq.~(\ref{kinetic})
and the (custodial-isospin violating) mass 
higher-dimensional operator
in Eq.~(\ref{mass})
translate into ``oblique'' parameters, $S$ and $T$
\cite{pt}, respectively:
\begin{eqnarray}
S & = & \frac{s}{ 2 \pi } \nonumber \\
T & = & \frac{ t }{ 8 \pi e^2 },
\end{eqnarray}
while the 
fermionic operators 
in Eq.~(\ref{fermionoperator}) are considered ``non-oblique''. 
However,
for a special form of fermion-{\em Higgs} higher-dimensional operators
in Eq.~(\ref{fermionoperator}), these can be 
field-redefined into {\em purely} oblique effects as we now discuss.
In the particular RS models studied in references 
\cite{csaki, carena1}, with all fermions
on the IR brane, the equivalent of our field redefinition was achieved by 
setting the gauge boson wavefunction to be unity on the IR brane. 
Reference \cite{carena2} studied {\em bulk} fermions and used analogous field 
redefinitions.

This special form is 
\begin{eqnarray}
x & = & a g^2 \nonumber \\
y & = & a g^{\prime \; 2} Y Y^H
\end{eqnarray}
for {\em all} fermions, 
where $Y = Q_{ em } - \tau^3_{ L }$
denotes the hypercharge of the fermion
and $Y_H$ is the hypercharge of the Higgs. 
Setting Higgs to its vev in the fermion-Higgs operator 
in Eq.~(\ref{fermionoperator})
induces non-canonical couplings of fermions to gauge bosons. However, doing
the following redefinition of gauge fields renders the fermionic couplings
to gauge bosons canonical\footnote{This can be extended (non-linearly)
into a manifestly $SU(2)_L \times U(1)_Y$-invariant redefinition.}:
\begin{eqnarray}
W_3 & \rightarrow & W_3 \left( 1 - g^2 \frac{a}{ 64 \pi^2 } \right) +
B g g^{ \prime } \frac{a}{ 64 \pi^2 } \nonumber \\ 
W^{\pm} & \rightarrow & W^{\pm} \left( 1 - g^2 \frac{a}{ 64 \pi^2 } \right) 
\nonumber \\  
B & \rightarrow & B\left( 1 - g^{ \prime \; 2 } \frac{a}{ 64 \pi^2 } \right) +
W_3 g g^{ \prime } \frac{a}{ 64 \pi^2 }. 
\label{redef}
\end{eqnarray}
This redefinition when substituted in 
SM gauge kinetic terms 
induces shifts in $s$ and $t$
i.e., purely oblique effects,
so that we now have:  
\begin{eqnarray}
S & = & \frac{s}{ 2 \pi } + \frac{a}{ 2 \pi } \nonumber \\
T & = & \frac{t}{ 8 \pi e^2 } + \frac{ a g^{ \prime \; 2 } }{ 8 \pi e^2 }
\label{totalST}
\end{eqnarray}
As we will show, in our model, the fermion-Higgs 
operators have the special form only for the {\em light} fermions
(and {\em right}-handed bottom), {\em not} for the top and
{\em left}-handed 
bottom quark so that as far as the precision electroweak fit is concerned, 
couplings of $b_L$ have to be considered separately. Hence, we will focus on
$S$ and $T$ parameters and $Z \rightarrow \bar{b}_L b_L$ in this paper.

In the following, 
we will integrate out Kaluza-Klein (KK) ({\em heavy}) modes of gauge/fermion fields in the
RS1 model and compute the
resulting dimension-$6$ operators
for the (light) 
{\em zero}-modes (which correspond to the 
SM fields
in the above Lagrangian).
There are even higher-dimensional operators whose effect on electroweak
precision observables can be considered. Naively, these are further
suppressed by $\sim g^2 v^2 / m_{KK}^2$, where $m_{KK}$ is the KK 
mass scale. In the present scenario, these are also $k \pi r_c$-enhanced.
Therefore, we are careful in what follows to consider KK mass scales such 
that even with this enhancement, these operators are suppressed relative to the
dimension-$6$ operators. This justifies using a Higgs vev 
insertion approximation
within gauge and fermions propagators in what follows.

\section{Tree-level $S$ and $T$ contributions from gauge-Higgs sector}
\label{STgauge}

\subsection{Contribution to $T$}
\label{Tgauge}

A powerful aspect of our model is that the bulk right-handed gauge symmetry enforces 
custodial isospin.  We will see that the gauge sector does not make a logarithmically 
enhanced contribution to the $T$ parameter, and that the $S$ 
parameter is log enhanced, but suppressed by $(v z_v)^2$ relative to $T$.  

In the effective theory (section {\ref{formalism}), 
the operator $t  | H^{ \dagger } D_\mu H |^2 / (4 \pi v)^2$ 
contributes only to
the $W^3$ mass at order $v^4$.
Consequently, the coefficient $t$ measures the amount of isospin breaking. In terms of vacuum polarizations, the effective theory contains a modified quadratic term,
\be
\mathcal{L}_{\mathrm{eff}} \supset 
g^2 \Big(\frac{v^2}{8}+ \frac{\Pi_{aa}(0)}{2} \, \Big) 
W_{\mu}^{a \,(0)} W^{a \, (0) \mu}.
\label{massterm}
\ee 
$\Pi_{ab}(q)$ is 
polarization
from integrating out tree level insertions of gauge KK modes. 
Thus, from Eqs.~(\ref{mass}) and (\ref{massterm}), we see that the coefficient of the operator $| H^{ \dagger }  D_\mu H |^2 / (4 \pi v)^2$ is
\be
t= - \frac{ 128 \pi^2 }{v^2} \; \left( \Pi_{33} (0) - \Pi_{11} (0) \right).
\ee

\begin{figure}[t]
\centering
\epsfig{file=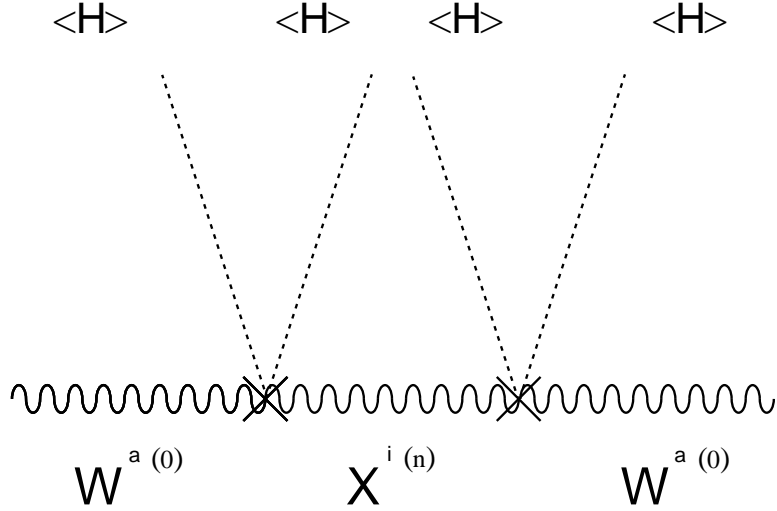, width=.6\linewidth}
\caption{Contributions 
to $S$ and $T$ from gauge-Higgs sector at tree-level.  
As in appendix \ref{KKsum5Dprop}, $X^i$ is any field coupling to the $W^a$'s.}
\label{Ttree}
\end{figure}

In the gauge sector, oblique corrections to the electroweak observables come from integrating out KK towers which couple
to left handed zero modes (Fig.~\ref{Ttree}).  We will find it convenient to convert sums over KK propagators 
(eigenfunctions) to five dimensional propagators (Green's functions)
(see appendix \ref{5Dprop}), while leaving four 
dimensional fields on external lines (see appendix \ref{KKsum5Dprop} for details).
   
We use Eq.~(\ref{kktoG}) 
to calculate the contribution to $\Pi_{aa}(q)$ from Fig\ref{Ttree},
\be
\label{Adef}
i g^2 \Pi_{aa}(q) \eta^{\mu \nu} = \sum_{j} i \frac{v^4}{16} g^2 g^2_{5i} \left( 
G^{5D}_{q \, j}(z_v, z_v) -  G^{(0)}_{ q \, j } \right) \eta^{\mu \nu},
\ee 
where $g = g_5 / \sqrt{ \pi r_c }$ is the $4D$ or zero-mode
gauge coupling.
The sum over $j$ includes all fields which couple to the external $W^{a \, (0)}$.
$G^{ 5D }_{ q j }$ in  
(\ref{Adef}) 
is the IR brane to IR brane, 
five dimensional, mixed momentum-position space Green's function in Feynman gauge.  
We \emph{subtract} the massless pole, $G^{(0)}_{q \, j}(z_v, z_v)$, since the effective Lagrangian is obtained by integrating out {\em only}
heavy/KK modes at tree-level (graphs with internal zero-modes
are part of the Dyson resummation). 

In the charged sector, the $W^{1 \, (0)}$ mixes with its own KK modes and those 
of the $\widetilde{W}^{1 \, (n)}$:
\be
\label{A11}
\Pi_{11}(q)= \left( \frac{v^2}{4} \right)^2 \Big\{g^2_{5} \left( G_{q(++)}^{5D} (z_v, z_v) - 
G_{q(++)}^{(0)} \right) + \tilde{g}^2_{5} G_{q(-+)}^{5D} (z_v, z_v) \Big\}.
\ee
In the neutral sector, $W^{3 \, (0)}$ mixes with its own KK 
modes, the $B$, and those of the $Z^{\prime}$: 
\be
\label{A33}
\Pi_{33}(q)= \left( \frac{v^2}{4} \right)^2 \Big\{ (g^2_{5}+g^{\prime 2}_5)  \left( G_{q(++)}^{5D}
(z_v, z_v) - 
G_{q(++)}^{(0)} \right) + g^2_{ Z^{\prime} \; 5 } \cos ^4 \theta^{ \prime } 
G_{q(-+)}^{5D} (z_v, z_v)  \Big\}.
\ee
The $\Pi$'s contain IR brane to IR brane
propagators and we have used the $Z^{ \prime }$ charge of Higgs
(see Eq.~(\ref{QZprime})).

Using the propagators in the appendix \ref{KKsum5Dprop}, 
we find
\be
\label{A110}
\Pi_{11}(0) \approx   \frac{-(k \pi r_c) g^2 (v z_v)^2 }{8} \left\{ \; 1 -  
\frac{1}{k \pi r_c}  + \mathcal{O} \left(   \frac{1}{(k \pi r_c)^2} \right) \right\} 
\frac{v^2}{4} -  \frac{(k \pi r_c) \tilde{g}^2 (v z_v)^2}{8} \left( 1
- \frac{ \tilde{M}^2 }{ 4 k^2 } \right)  \frac{v^2}{4}, 
\ee
and
\begin{eqnarray}
\label{A330}
\Pi_{33}(0) \approx \frac{-(k \pi r_c)(g^2+g^{\prime 2})(v z_v)^2}{8} \left\{ \; 1 -  
\frac{1}{k \pi r_c}  + \mathcal{O} \left( \frac{1}{(k \pi r_c)^2} \right) \; \right\} 
\frac{v^2}{4} & - &  \nonumber \\
 \frac{(k \pi r_c)( g^2_{ Z^{\prime} } \cos^4 \theta^{ \prime } )
(v z_v)^2}{8} \frac{v^2}{4} &  & 
\end{eqnarray}
where $g^{ \prime } = g^{ \prime }_5 / \sqrt{ \pi r_c }$, $\tilde{g} =
\tilde{g}_5 / \sqrt{ \pi r_c }$
$g_{ Z^{\prime} } =  g_{ Z^{ \prime } \; 5 } / \sqrt{ \pi r_c }$.
In (\ref{A110}) and (\ref{A330}), we have also dropped terms at order $v^4$ that are 
$e^{ - k \pi r_c }$ suppressed relative to the leading terms above.  
Then,
\be
\Pi_{33}(0)-\Pi_{11}(0) \approx  \frac{(g^{\prime 2})(v z_v)^2}{8} \frac{v^2}{4}
- \frac{ \tilde{M}^2 }{ 4 k^2 }
\frac{ k \pi r_c }{8} \tilde{g}^2 \left( v z_v \right)^2 \frac{ v^2 }{4}, \quad \mathrm{and} 
\ee
\be
t = 16 \pi^2 v^2 z_v^2
\left( \frac{ \tilde{M}^2 }{ 4 k^2 } \frac{ \tilde{g}^2 }{4} k \pi r_c - 
\frac{ g^{ \prime \; 2} }{4}
\right).
\label{tRS}
\ee
Hence,
in scenario II,
the gauge sector \emph{does not} contribute a log enhanced piece to custodial $SU(2)$ 
breaking at this order 
as would be the case in the absence of $SU(2)_R$. 

\subsection{Contribution to S}
\label{Sgauge}

At tree-level, there are no Feynman diagrams that contribute to 
$s$ as defined in Eq. (\ref{kinetic}). As discussed at the end of section
\ref{formalism}, there may be even higher-dimensional operators which
can contribute to precision variables which we argued on general grounds
are small in our model. However, in the case of the $S$ parameter
since the dimension-$6$ contribution, $s$, vanishes, to be cautious, we will
calculate dimension-$8$ contribution of the type
depicted in Fig.\ref{Ttree}.

Since this is a tree-level calculation, there is no (kinetic) 
mixing involving
the photon (of course, there can{\em not} be any mass mixing even at loop
level), so the quantity
\be
\left( \Pi ^{\prime}_{33}(q)|_{q^2=0} - \Pi ^{\prime}_{3Q}(q)|_{q^2=0} \right) 
\approx  \Pi ^{\prime}_{33}(q)|_{q^2=0},
\ee
is related to $S$, 
with 
\be
\Pi ^{\prime}_{33}(q)|_{q^2=0} 
\approx  \frac{-(g^2+g^{\prime 2}+4 g^2_{Z^\prime H})(k \pi r_c)
(v z_v)^4}{256} + \mathcal{O}(v^4 z_v^4).
\label{Sdim8}
\ee
Eq. (\ref{Sdim8}) is equivalent to a shift in $S$ parameter,
$\Delta S \approx 16 \pi \Pi^{ \prime} _{33}(q)|_{q^2=0}$ \cite{pt}. 
As we will see, this contribution can be neglected.


The $\Pi$'s contribute to $S$ and $T$, but we postpone a detailed discussion of the 
model's prediction for electroweak precision measurements, since, as discussed 
in section \ref{formalism}
additional contributions to 
$S$ and 
$T$ arise from the rescaling of gauge bosons.  A careful treatment of the fermionic 
sector 
is necessary.

\section{Fermionic operators}

The coefficients of the operators
in Eq.~(\ref{fermionoperator}) get contributions from integrating out KK modes
of gauge fields (at tree-level) as shown in the Feynman
diagrams of Figs.~\ref{Stree} and \ref{compositeness}.
\begin{figure}[t]
\centering
\epsfig{file=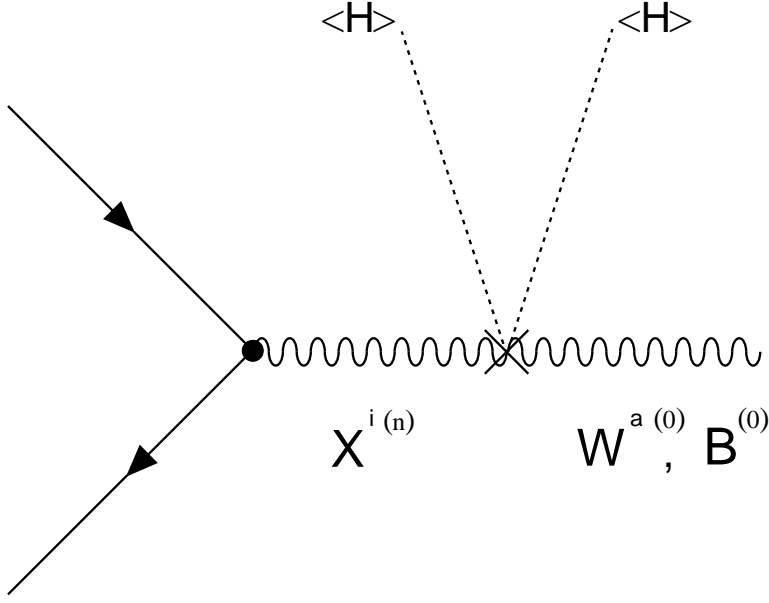,width=.6\linewidth}
\caption{Contribution to fermion-Higgs operator.
As in appendix \ref{KKsum5Dprop}, $X^i$ is any field coupling to the $W^a$'s.}
\label{Stree}
\end{figure}
\begin{figure}[t]
\centering
\epsfig{file=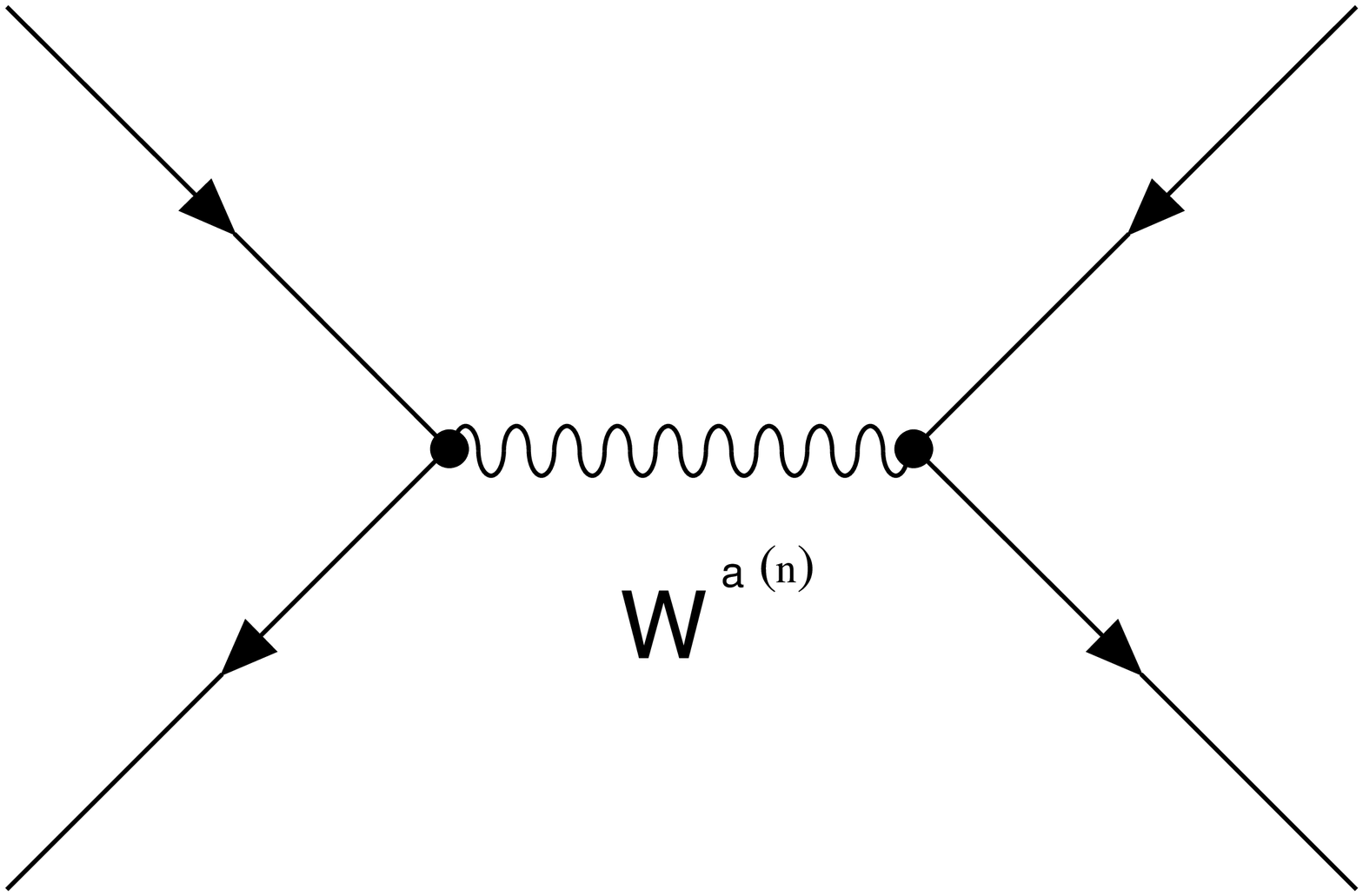,width=.6\linewidth}
\caption{Contribution to $\psi^4$.}
\label{compositeness}
\end{figure}
The Feynman diagram in Fig.~\ref{Stree}
is evaluated by 
integrating fermion zero-mode wavefunction with propagator from $z$ to TeV brane
(including metric/f\"unfbein factors). This 
gives the fermion-Higgs higher-dimensional operator in
Eq.~(\ref{fermionoperator})
with
coefficients (up to $O \left( z_h^2 / z_v^2 \right)$)  as follows\footnote{See 
also references \cite{shafi, hewett2, burdman, carena2} 
for discussion of this effect in a different
language.}. From exchange of KK modes of $W_L$, we get
(see appendix \ref{appfermion} for details)
\begin{eqnarray}
x (c) & = & 16 \pi^2 v^2 g_5^2 \int d z \sqrt{G} \; \frac{z}{z_h} \;
\chi_0^ 2 (c, z) \left( G^{ 5D }_{q =0  
(++) } (z, z_v) - G_{ q (++) } ^{ (0) } \right) \nonumber \\ 
 & = & g^2 \left( 16 \pi^2 v^2 z_v^2 \right) \big[ \frac{1}{4}
\left( 1 - \frac{1}{k \pi r_c} \right) + \frac{1}{ 1 - e^{ k \pi r_c 
(2c-1) } }
\frac{1-2c}{3-2c} \left( - \frac{ k \pi r_c }{2} + \frac{ 5-2 c}{ 4 ( 3-2c ) }
\right) \big], \nonumber \\
\label{coeffw}
\end{eqnarray}
From exchange of KK modes of hypercharge and $Z^{ \prime }$ 
%
\begin{eqnarray}
y (c) & = & 
16 \pi^2 v^2 g^{\prime \;2}_5 Y Y^H \int d z \sqrt{G} \; \frac{z}{z_h} \; 
\chi_0^2 (c, z) \left( 
G^{ 5D }_{q = 0
(++) } (z, z_v) - G^{ (0) }_{ q (++) } \right)  \nonumber \\
 & & + 16 \pi^2 v^2 g^2_{ Z^{\prime} \; 5 } Q_{ Z^{\prime} } Q^H_{ Z^{\prime} } 
\int d z \sqrt{G} \; \frac{z}{z_h} \;  
\chi_0^2 (c, z) \left( G^{ 5D }_{q = 0
(++) } - G_q^{ (0) } \right) (z, z_v) \nonumber \\
 & = & g^{\prime \; 2} \left( 16 \pi^2 v^2 z_v^2 \right) Y Y^H \big[ \frac{1}{4}
\left( 1 - \frac{1}{k \pi r_c} \right) + \frac{1}{ 1 - e^{ k \pi r_c 
(2c-1) } }
\frac{1-2c}{3-2c} \left( - \frac{ k \pi r_c }{2} + \frac{ 5-2 c}{ 4 ( 3-2c ) }
\right) \big] \nonumber \\
 & & + g^2_{ Z^{ \prime } } Q_{ Z^{\prime} } Q^H_{ Z^{\prime} }
\left( 16 \pi^2 v^2 z_v^2 \right) \frac{1}{ 1 - e^{ k \pi r_c 
(2c-1) } }
\frac{1-2c}{3-2c} \left( - \frac{ k \pi r_c }{2} \right)
\nonumber \\
\label{coeffy}
\end{eqnarray}
%
The factor of $z/z_h$ (inside
$z$ integral) in first line is
from f\"unfbein.
$Q_{ Z^{\prime} }$
and $Q^{H}_{ Z^{\prime} }$ 
denotes the $Z^{ \prime }$
charge of fermion and Higgs (see Eq.~(\ref{QZprime}). 

A similar computation of the Feynman diagram in
Fig.~\ref{compositeness}
gives coefficient of ``compositeness''
operator in Eq.~(\ref{fermionoperator}), 
$\bar{ \psi } \tau^a \psi
\bar{ \psi ^{ \prime} } \tau_a \psi^{ \prime }$ (from exchange of KK modes of $W$): 
\begin{eqnarray}
V ( c, c^{ \prime } ) & = & 16 \pi^2 v^2 g_5^2 \int d z
\sqrt{ G (z) } \; \frac{z}{z_h} \; 
\psi^{ (0) \; 2}  ( c, z ) 
\int d z^{ \prime } \sqrt{ G ( z^{ \prime } ) } \; \frac{ z^{ \prime } }{z_h} \; 
\psi^{ (0) \; 2 } ( c^{ \prime }, z^{ \prime } ) 
\nonumber \\
& & \times
\left( G^{ 5D }_{q =0
(++) } (u, v) - G^{ (0) }_ q \right)
\nonumber \\
\end{eqnarray}
We obtain $V \approx g^2 \; 16 \pi^2 v^2 z_v^2 / ( 4 k \pi r_c )$ 
for $c, c^{ \prime} > 1/2 + \epsilon$ (as applicable
to light fermions): this coefficient is
negligible for $z_v^{-1} \stackrel{>}{\sim}$ TeV 
and similarly for exchange of hypercharge or $Z^{ \prime }$
KK modes\footnote{Coefficient of the operator
(light fermion)$^2$ (top or {\em left}-handed bottom)$^2$
will be larger since
$c$ for top quark or {\em left-handed} bottom $< 1/2$ (see section
\ref{topbottom}), but it plays no role in
fit to precision electroweak data,
although it will affect, say, $e^+ e^-
\rightarrow \bar{b}_L b_L$ at high-energy colliders.}. 
See also references \cite{gp, hewett1}.

\subsection{Light fermions}
\label{lightfermion}

If light fermions are at $c > 1/2 + \epsilon$ (such that
$e^{ k \pi r_c \epsilon } \gg 1$: $\epsilon \stackrel{>}{\sim} 0.1$ suffices)
in order to address flavor (as mentioned before \cite{gp, huber}), then 
second term proportional to $Y$
in Eq.~(\ref{coeffy}) can be neglected. Also, 
KK modes of
$Z^{\prime}$ 
couple
very weakly to light fermions and so there are no operators proportional 
to
$Z^{\prime}$ charge (i.e., last term in Eq.~(\ref{coeffy})).
Thus, coefficients $x$ and $y$ of these fermion-Higgs operators 
are of the 
special form
discussed in section \ref{formalism}
with 
\begin{equation}
a \approx 4 \pi^2 v^2 z_v^2.
\end{equation}
Hence, the $S$ parameter in our model is given by
(see Eq.~(\ref{totalST}))
\begin{eqnarray}
S & \approx & 2 \pi v^2 z_v^2  \nonumber \\
 & \approx & 2 \pi v^2 \frac{6}{ m^{ (1) \; 2 }_{ gauge} },
\label{SRS}
\end{eqnarray}
where 
$m^{ (1) }_{ gauge}$ is the mass of the lightest KK mode of gauge boson  
(see Eq.~(\ref{lightestgaugeKK})).
Here we have 
neglected $S$ from gauge-Higgs sector since it is of higher order
in $z_v v$
(see section \ref{Sgauge}).

There is an interesting possibility that we will not pursue here
where contribution to $S$ parameter arising from the fermion-Higgs operators
in Eq.~(\ref{fermionoperator}) is suppressed completely for $c = 1/2$
as can be seen from Eqs.~(\ref{coeffw}) and (\ref{coeffy}). 
However, in order to fit observed light fermion masses with
$c = 1/2$, we would have to introduce very small dimensionless
numbers into our fundamental theory. While this is radiatively stable,
it goes against the general philosophy adopted in this paper.

Similarly, the $T$ parameter is given 
by (using Eqs.~(\ref{totalST}) and (\ref{tRS}))  
\begin{eqnarray}
T & \approx & \big[ \frac{ \pi }{2} \frac{ \tilde{g}^2 }{ e^2 } v^2 z_v^2 k \pi r_c 
\big]
\frac{ \tilde{M}^2 }{ 4 k^2 } \nonumber \\
 & \approx & \big[ 3 \pi \; k \pi r_c 
\frac{ v^2 }{ m^{ (1) \; 2 }_{ gauge } } 
\frac{ \tilde{g}^2 }{ e^2 } \big] \frac{ \tilde{M}^2 }{ 4 k^2 },
\label{Tbulkbreaking}
\end{eqnarray}
where
the quantity in
$[...]$'s is the $T$ parameter (assuming $\tilde{g} = g^{ \prime }$) 
that would result if we repeated our analysis with {\em purely}
SM gauge group in bulk rather than our present
extended gauge sector.

\subsection{Top and bottom}
\label{topbottom}

We can obtain the $4D$ Yukawa coupling
$\lambda$ in terms of the $5D$ Yukawa coupling $\lambda_{5}$ (see, for example, \cite{gp}):
\begin{eqnarray}
\lambda & = & \lambda_{5} k 
\frac{ \sqrt{| (1-2c_L) (1-2c_R)|} }
{ \left( 1 - e^{ k \pi r_c ( 2 c_L -1 ) } \right) 
\left( 1 - e^{ k \pi r_c ( 2 c_R -1 ) } \right)}
\label{lambda4D}
\end{eqnarray}

The quick argument for choosing $c_{L, R} < 1/2$ for top quark
is that, 
for $c_L$ (or $c_R$) $> 1/2$, the $4D$ Yukawa coupling is
(exponentially) suppressed (see Eq.~(\ref{lambda4D})) 
and hence we consider $c_{L, R} < 1/2$
for top quark to obtain $\lambda _t \sim 1$. 


For $c_{ L, R } < 1/2 - \epsilon$, we get
\begin{eqnarray}
\lambda_{t} & \approx & \lambda_{ t \; 5 } k 
\sqrt{ (1-2c_L) (1-2c_R) } 
\label{lambda4Dtop}
\end{eqnarray}

Since coefficient
in Eq.~(\ref{coeffy}) is different for $b_L$ than for light fermions,
the effect on coupling of $b_L$ to $Z$
arising from the operators in Eq.~(\ref{fermionoperator})
can{\em not} be redefined into $S$ (see
discussion
in section \ref{formalism})
and must be treated separately:
\begin{eqnarray}
\frac{ \delta \left( g^{ b_L}_Z \right) }{ g^{ b_L}_Z }
& \approx & m_Z^2 z_v^2 
\left( \big[ 1 + \frac{ g^2_{ Z^{ \prime } } Q_{ Z^{ \prime} } Q^H_{ Z^{ \prime } } }
{ g_Z^2 Q_Z Q^H_Z } 
\big]
\frac{1-2c}{3-2c} \left( - \frac{ k \pi r_c }{2} + \frac{ 5-2 c}{ 4 ( 3-2c ) } \right)
\right. \nonumber \\
 & \approx & \frac{ m_Z^2 } { \left( 0.4 m_{ gauge }^{ (1) } \right) ^2 } 
\left( - \frac{ k \pi r_c }{2} 
+ O(1) \right) \; 0.9 \; \frac{1-2c}{3-2c}
\label{Zbb}
\end{eqnarray}
using 
the $Z^{ \prime }$ charges of Higgs and
$b_L$. 
Here,
$Q_Z = \tau^3_{L} - Q_{em} \sin ^2 \theta _W$ 
and $Q_Z^H$ are $Z$ charges of $b_L$ and Higgs.


To obtain $m_b \ll m_t$ with{\em out} hierarchy in $5D$ Yukawa coupling
($c_L$ is {\em same} for top and bottom), 
we choose $c$ for $b_R$ $> 1/2$.

\section{$T$ at loop level in Scenario II}
\label{Tloop}

An interesting case to consider is when $SU(2)_R$ is unbroken in the bulk 
(our Scenario II), 
$\tilde{M} = 0$. In that case, remarkably, bulk custodial isospin
symmetry forces loop contributions to $T$ parameter to be
UV finite (and hence calculable) and 
these are the dominant contribution to $T$.
This is because 
contribution to $T$ requires both electroweak symmetry breaking on IR brane and $SU(2)_R$
breaking which is localized on UV brane.
For remainder of this section, we will consider this case.

Because custodial-isospin violation is due to breaking of $SU(2)_R$ by boundary
condition on Planck brane, there is 
{\em no} zero-mode for $\tilde{W}^{ \pm }$
and KK spectrum
is different for $\tilde{W}^{ \pm }$ and $\tilde{W}^3$
(see appendix \ref{KKTeVcoupling}). Similarly,
there is {\em no}
zero-mode for $d^{ \prime }_R$ and KK spectrum
can be different for $u_R$ and $d^{ \prime }_R$ (as can be seen from appendices
\ref{spectrum++} and \ref{spectrum-+}).
Hence, 
loop diagrams will have to involve {\em right}-handed $\tilde{W}$ and/or 
$t$ (and other fermions) 
in order to give $T$.

An example of a Feynman diagram with $\tilde{W}$ zero and KK modes, but
with{\em out} fermions is shown in Fig.~\ref{Tgaugeloop}. 
This diagram with
$\tilde{W}^3$ {\em zero}-mode 
gives 
$W^{ \pm } W^{ \mp }$ mass term 
$\sim 
g^2 \frac{ g^{ \prime \; 2} g^2 2 k \pi r_c }{ 16 \pi ^2 }
\frac{ v^4 }{ m_{ gauge }^{ (1) \; 2} }$. 
A brief explanation is as follows: the 
quartic
$W$ vertex is $g^2$ since external legs are {\em zero}-modes and 
each Higgs vertex gives $g^{ \prime } g \sqrt{2 k \pi r_c}$, where 
$g^{ \prime }$
is due to $\tilde{W}^3$ (or hypercharge) {\em zero}-mode propagator
and $g \sqrt{2 k \pi r_c}$ is  
coupling of 
$W$ KK mode to Higgs on TeV brane (see appendix \ref{KKTeVcoupling}).
This diagram does not
give $W^3 W^3$ mass term since there is no quartic $W^3$ coupling. 
We can
estimate contribution with {\em KK} modes of $\tilde{W}^{ \pm , 3 }$ as follows: 
KK modes of $\tilde{W}^{ \pm }$ and
$\tilde{W}^3$ are split at $O \left( 1 / \big[ k \pi r_c \big] \right)$
(see appendix \ref{KKTeVcoupling}), whereas their coupling
to Higgs is enhanced by $\sim \sqrt{ k \pi r_c }$ (compared to $\tilde{W}$ zero-mode)
to so that
KK contribution (to both $W^{ \pm }W^{ \mp }$ and $W_3 W_3$ mass terms)
is comparable to that of $\tilde{W}^3$ zero-mode.

\begin{figure}[t]
\centering
\epsfig{file=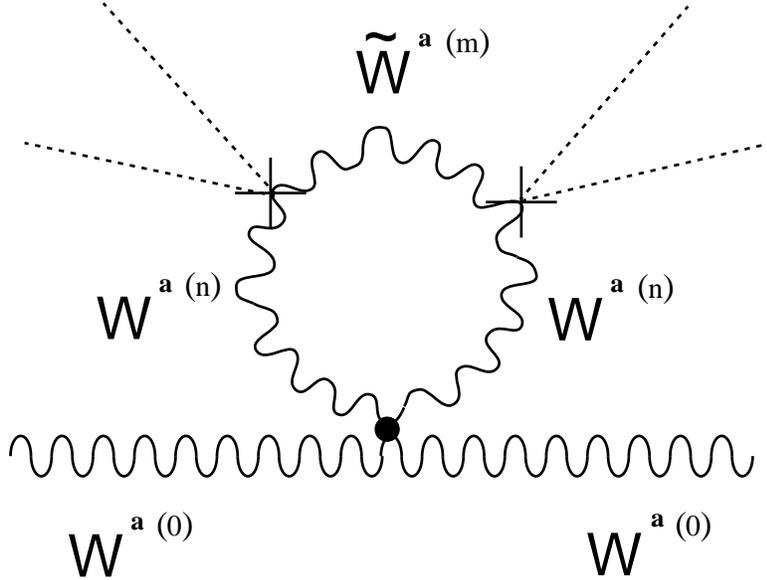,width=.6\linewidth}
\caption{Contribution to $T$ from gauge loop}
\label{Tgaugeloop}
\end{figure}

The Feynman diagrams with $t_R$ (and other {\em right}-handed fermions)
are shown in Figs.~\ref{Ttoploop1} and \ref{Ttoploop2}. 

Let us estimate the
Feynman diagram with{\em out} Yukawa insertion (Fig.~\ref{Ttoploop1}). The
contribution of $t_R^{ (0) }$ gives
$\Pi \sim \left( \frac{1}{2} \right)^2  
\frac{3}{ 16 \pi ^2 } \tilde{g}^4 \left( 
\sqrt{ 2 k \pi r_c } \right)^4 
\frac{ \left( v / 2 \right)^4 }{ m_{ gauge }^{ (1) \; 2 } }$, where
the factors of $1/2$ are from quantum number of $t_R$ and Higgs and the 
coupling of $t_R$ zero-mode (and Higgs) to $\tilde{W}^3$ {\em KK} mode 
is enhanced by $\approx \sqrt{ 2 k \pi r_c }$
(compared to $\tilde{W}^3$ zero-mode)
since $t_R^{ (0) }$ is localized near TeV brane.
The fractional mass splitting of {\em KK} 
modes of $t_R$ and $b^{ \prime }_R$ can be $O(1)$ 
depending on $c_R$ (see appendices
\ref{spectrum++} and \ref{spectrum-+}), 
while their couplings to KK modes of $\tilde{W}$
are almost same as that of $t_R^{ (0) }$ since they are also localized
near TeV brane. Hence, contribution of KK modes can be comparable to that of $t_R^{ (0) }$.

As we will see,  
the
Feynman diagram 
involving insertions of
{\em top} Yukawa coupling, Fig.~\ref{Ttoploop2} dominates over
the diagram 
with{\em out} top Yukawa insertion
(Fig.~\ref{Ttoploop1}) (and diagrams with other fermions) and also over 
the diagram with{\em out} fermions
(Fig.\ref{Tgaugeloop}). So, we 
concentrate on the diagram in Fig.\ref{Ttoploop2}.

\begin{figure}[t]
\centering
\epsfig{file=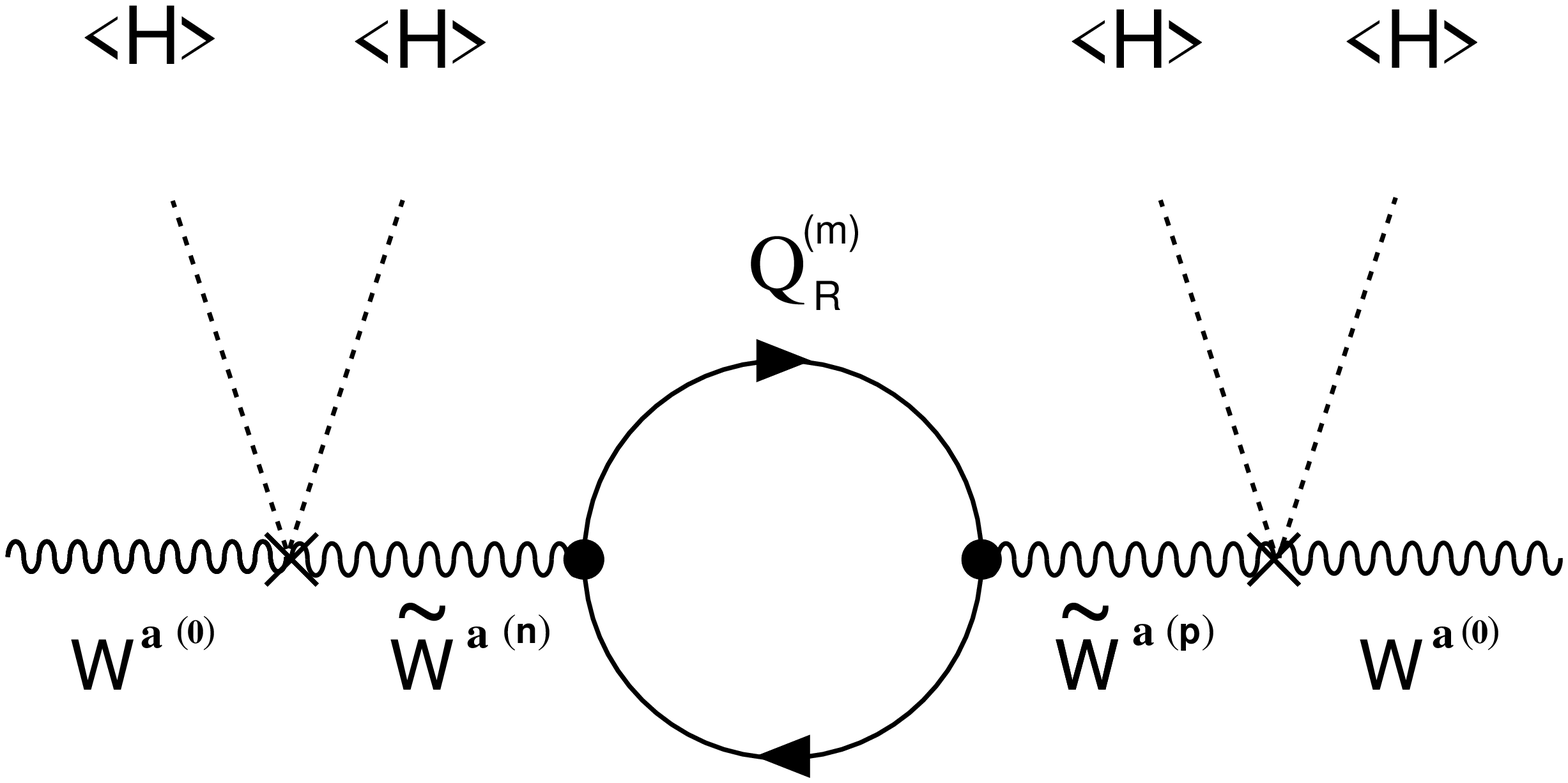,width=.6\linewidth}
\caption{Contribution to $T$ from top loop with{\em out} Yukawa insertion.}
\label{Ttoploop1}
\end{figure}
\begin{figure}[t]
\centering
\epsfig{file=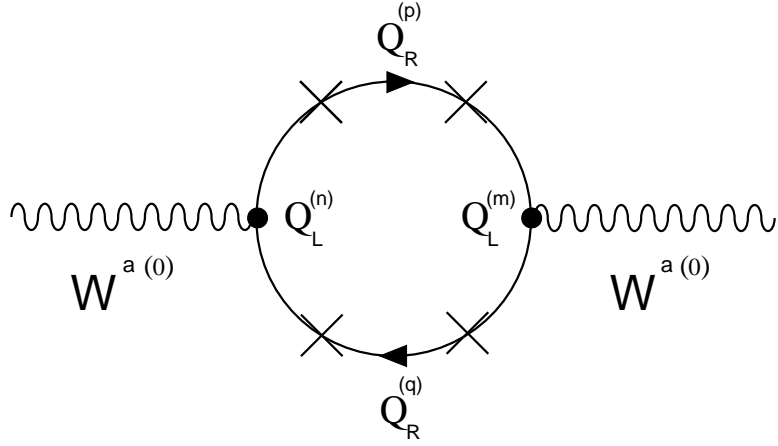,width=.6\linewidth}
\caption{Contribution to $T$ from top loop with Yukawa insertion.
The ``$\times$'' denotes a $t_L^{ (m) } t_R^{ (n) }$ or
$b^{ (m) }_L b^{ \prime \; (n) }_R$ ``mass insertion''.}
\label{Ttoploop2}
\end{figure}

\subsection{$T$ from top quark KK modes}
\label{Ttop}

To calculate the diagram in Fig.~\ref{Ttoploop2},
we need (a) spectrum of KK modes of {\em left}-handed top and bottom
and also $t_R$ and its $SU(2)_R$ partner, $b^{ \prime }_R$ ({\em not} the physical $b_R$)
and (b) their couplings to Higgs (or the $t_L^{ (m) } t_R^{ (n) }$
and $b^{ (m) }_L b^{ \prime \; (n) }_R$ ``mass insertions'').
The spectrum and couplings to Higgs
are calculated in the appendices \ref{spectrum++} and \ref{spectrum-+} to which
the reader is referred for details.
We find it
{\em not} 
convenient to convert sum over KK modes into propagators (unlike before) and so
we will work directly in terms of KK modes.

We begin with $c_R < -1/2 - \epsilon$, where $\epsilon
\stackrel{>}{\sim} 0.1$. 
In this case,
we get a ``very light'' (much lighter than
$z_v^{ -1 }$) $b^{ \prime }_R$ 
mode. This is
ruled out experimentally. 

For a slightly larger $c_R$, namely, $c = -1/2 + \epsilon$ (with $\epsilon \sim 0.1$),
we can show that 
the lightest 
$b^{\prime}_R$ 
mode is {\em not}
lighter than
$z_v^{ -1 }$
(and hence {\em not} ruled out experimentally
unlike before), but its mass is smaller
than that of the lightest $(t, b)_L$ and $t_R$ KK mode.
Other modes of $b^{\prime}_R$ are almost degenerate 
with KK modes of
$t_R$. 
Also, because $c_R \approx 1/2$,
mass 
insertions in Fig.~\ref{Ttoploop2} involving $t_R$ and $b^{ \prime }_R$ 
{\em KK} modes are 
the {\em same} as those involving the $t_R$ {\em zero}-mode. 
One can show that this results in small $T$.

Next, we consider $c_R \sim 0$.
It is clear that
{\em all} modes of $t_R$ and $b^{ \prime }_R$ need to be considered
since KK spectrum is different for $t_R$ and $b^{ \prime }_R$.
The $t_L^{ ( n \neq 0 ) } t_R^{ (0) }$ mass insertion is 
$m_t f (c_L)$, whereas the $t_L^{( m \neq 0 )} t_R^{( n \neq 0 )}$ 
and $b_L^{ ( m \neq 0 ) } b^{ \prime \; (n) }_R$
mass insertions are $m_t f (c_L) f (c_R)$, where $f (c)$ is given in Eq.~(\ref{f}). 
For any given {\em KK} modes of
$t_L$ and 
$t_R$ 
in the loop, 
the diagram in Fig.~\ref{Ttoploop2} gives
(with {\em four} insertions
of $t_L^ { (m \neq 0) } t_R^{ (n \neq 0) }$) 
%
\begin{eqnarray}
\Pi _{ 33 } (0) (m,n,p, q) 
& \approx & \frac{-3}{ 64 \pi^2 } 
\frac{ f (c_L)^4 f (c_R)^4 m_t^4 }{ m_{ t^{ (m) }_L }^2 } 
\nonumber \\ & & \times
\int_0^{ \infty } dx 
\frac{ x^2 \big[ x^2 + 2 \left( 1 + 
\frac{ m_{ t^{ (n) }_L }^2 }{ m_{ t^{ (m) }_L }^2 } \right) x + 
\frac{ m_{ t^{ (n) }_L }^2 }{ m_{ t^{ (m) }_L }^2 }
\big] }{ \left( x + 1 \right)^2 
\left( x +\frac{ m_{ t^{ (n) }_L }^2 }{ m_{ t^{ (m) }_L }^2 } \right)^2 
\left( x + \frac{ m_{ t_R^{ (p) } }^2 }{ m_{ t^{ (m) }_L }^2 } \right) 
\left( x + \frac{ m^2_{ t 
_R ^{ (q) } } }{ m_{ t^{ (m) }_L }^2 } \right) } 
\nonumber \\
\label{Pi}
\end{eqnarray}
There are similar contributions to $\Pi_{ 33 }$ involving
{\em KK} modes of $b_L$ and $b^{ \prime }_R$ and to
$\Pi_{ 11 }$ involving {\em KK} modes of $(t,b)_L$ and $t_R$, $b^{ \prime }_R$. 
For a $t_R$ {\em zero}-mode in the loop, we set $f (c_R )$ to $1$. 
The spectrum is not very sensitive to $c_L$ and $c_R$, so we use 
spectrum for $c_L = 1/2$
since we will choose $c_L \sim 0.4$, i.e., close to $1/2$
in section \ref{fit}
and $c_R = 0$: 
\begin{eqnarray}
m_{ t_R^{ (n) } } z_v & \approx \hbox{zeroes of} \; J_{ \frac{1}{2} } 
& = n \pi \nonumber \\ 
m_{ b^{ \prime \; (n) }_R } z_v & \approx \hbox{zeroes of} \; J_{ - \frac{1}{2} } 
& = \left( n - \frac{1}{2} \right) \pi
\end{eqnarray}
and
\begin{eqnarray}
m_{ t_L^{ (n) }, \; b_L^{ (n) } } z_v & \approx & \hbox{zeroes of} \; J_0 \nonumber \\
 & \approx &
\left( n - \frac{1}{4} \right) \pi 
\end{eqnarray}
$f ( c_L )$ in Eq.~(\ref{Pi})
is sensitive to $c_L$ (see Eq.~(\ref{f})) since we will choose
$c_L \sim 0.4$ (i.e., close to $1/2$) in section \ref{fit}, but the
dependence on $f ( c_L )$
is analytic. Whereas,
$f (c_R)$ is {\em not} very sensitive to $c_R$ since $c_R \sim 0$ (see Eq.~(\ref{f}))
and so we
set $f (c_R) = f(0) = \sqrt{2}$ in Eq.~(\ref{Pi}).

A numerical
evaluation of the
loop integral and the sum over {\em KK} modes 
of $(t,b)_L$ (i.e., $m, n$) and {\em all} modes of $t_R$, $b_R^{ \prime }$ 
(i.e., $p, q$) 
in 
Eq.~(\ref{Pi}) gives
(as mentioned earlier, the sum over modes converges 
because of
bulk $SU(2)_R$ gauge symmetry) 
\begin{eqnarray}
T_{ \hbox{ KK top } } 
& = & \frac{ 4 \pi }{ \sin^2 \theta _W 
\cos^2 \theta _W M^2_Z } \big[ \Pi _{11} (0) -
\Pi _{33} (0) \big] 
\nonumber \\
& \approx & T_{ \hbox{ SM top } } \left( 
\frac{ f^2 (c_L ) m_t }{ m_{ t ^{ (1) }_L } 
} \right)^2 0.7,
\label{TtopKK}
\end{eqnarray}
where $0.7$ is from (numerical) integration and KK sum,
$m_{ t ^{ (1) }_L }$ is the mass of the lightest $(t,b)_L$ KK mode
and
$T_{ \hbox{ SM top } } = \frac{3}{ 16 \pi \sin ^2 _{ \theta _W }
\cos ^2 _{ \theta _W } } \frac{ m_t^2 }{ m_Z^2 } \approx 1.2$.

If we replace KK modes of {\em left}-handed top (or bottom)
by {\em zero}-mode in this Feynman diagram, then we lose a factor of
$f (c_L) \gg 1$ and so such effects are sub-leading.
Of course, Fig.~\ref{Ttoploop2} will {\em only} zero-modes in the loop
is the SM top quark
contribution to $T$. 

Now, we can see why $T$ from other (light) fermion KK modes 
is smaller than that 
from top quark.
The 
diagram in Fig.~\ref{Ttoploop2}
with {\em zero}-mode of, say, $u_R$
is smaller than in the case of top quark 
(even if $\lambda _{ 5 }$ is comparable for all 
fermions) due to smaller
mass insertion which, in turn, is due to larger $f (c_R)$ for light fermions.
Whereas, for KK modes of $u_R$, mass insertions 
can be {\em comparable} to that in the case of top quark
(if $\lambda _{ 5 }$ is comparable for all 
fermions), but the non-degeneracy of $u_R$ and $d^{ \prime }_R$ KK modes is 
smaller than in the case of top quark, again due 
to larger $f (c_R)$ for light fermions.
Hence, the
contribution of KK modes
of $u_R$ is also smaller than in the case of top quark.

\section{Electroweak fit}
\label{fit}

We now put the pieces together and fit the precision electroweak data.

\subsection{$Z \rightarrow \bar{b}_L b_L$}

From Eq.~(\ref{Zbb}), we see that, for $c_L 
\stackrel{>}{\sim} 0.3$ 
and $m_{ gauge }^{ (1) } 
\stackrel{<}{\sim} 4$ TeV,
the shift in the 
coupling of $b_L$ to $Z$ is $\stackrel{<}{\sim} 1 \%$
which is allowed by precision electroweak data. 
If $c_L \stackrel{<}{\sim} 0.3$, then $m_{ gauge }^{ (1) } \gg$ few TeV to be
consistent with $Z \rightarrow \bar{b}_L b_L$, a case not of interest to us here.
Hence, we choose $c_L \stackrel{>}{\sim} 0.3$.

\subsection{$S$ and required $T_{RS}$}

We choose 
$c > 1/2$ for light fermions
in order to address flavor
issues: hierarchy of fermion masses and suppression of FCNC's 
as
mentioned before \cite{gp, huber}. From Eq.~(\ref{SRS}), this gives
%
\begin{eqnarray}
S_{RS} & \sim & 0.2 
\; \; \hbox{for} \; m^{ (1) }_{ gauge } \sim 3-4 \; \hbox{TeV}.
\end{eqnarray}
%
A smaller $m^{ (1) }_{ gauge }$ will give too large
$S$ (unless we choose $c = 1/2$ for light fermions as we discussed in section
\ref{lightfermion}) such that we cannot fit data 
(independent of $T$)
So, we will
not consider $m^{ (1) }_{ gauge } < 3$ TeV.

A heavy Higgs, say, $m_H \sim 500$ GeV  
gives \cite{pdg}
\begin{eqnarray}
S_{ SM } & \sim  & +0.1 \nonumber \\
T_{ SM } & \sim & -0.15, 
\end{eqnarray}
where $S_{SM}$ is measured relative to SM with Higgs mass $100$ GeV.
Adding $S_{SM}$ to $S_{RS}
\sim + 0.2$ 
shows 
that 
\begin{eqnarray}
T^{required}_{RS} & \sim & + 0.35-0.5
\label{Treq1}
\end{eqnarray}
is required to get a reasonable fit to data \cite{pdg}.

For smaller Higgs mass, say, $\sim 200$ GeV we have \cite{pdg} 
\begin{eqnarray}
S_{SM} & \sim & +0.05 \nonumber \\
T_{SM} & \sim & -0.05
\end{eqnarray}
so that (adding it to $S_{RS}
\sim + 0.2$) shows  
\begin{eqnarray}
T^{required}_{RS} & \sim & +0.15-0.35 
\label{Treq2}
\end{eqnarray}
is required to fit data \cite{pdg}.

\subsection{Scenario I: $\tilde{M} \neq 0$}

From Eq.~(\ref{Tbulkbreaking}), we see that
the {\em tree}-level value of $T$ 
({\em after} fermion-Higgs operators are transformed into $S$ as discussed in section 
\ref{formalism})
is controlled by $\tilde{M} / k$.
The fractional splitting of 
the KK masses of $\tilde{W}^{ \pm }$
and $\tilde{W}^3$ is given by
$\sim \tilde{M}^2 / \left( 4 k^2 \right)$ (see appendix \ref{KKTeVcoupling})
so that,  
assuming that the fractional mass splitting $< 1/2$ to warrant approximation
of small custodial-isospin breaking
\begin{eqnarray}
T_{RS} &
\sim & 2 \times \frac{ \tilde{M}^2 }{ 4 k^2 }
\; \hbox{for} \; m^{ (1) }_{ gauge } \sim 3-4 \; \hbox{TeV}
\nonumber \\
 & \sim & 0-1,
\label{Trange1}
\end{eqnarray}
where we have assumed $\tilde{g} \sim g^{ \prime }$.
From Eqs.~(\ref{Trange1}), (\ref{Treq1}) and (\ref{Treq2}), 
our model can fit the data for both light and heavy Higgs
using control parameter $\tilde{M}/k$ for a sizable portion of its range.

\subsection{Scenario II: $\tilde{M} = 0$}


In the absence of bulk $SU(2)_R$ breaking, 
it is interesting to see if the required $T$ can be generated 
by radiative effects. We saw in
section \ref{Tloop} that $T$ from top loop in Fig.~\ref{Ttoploop2} dominates
and depends on $c_L$ and $c_R$ for top quark.
$T$ is 
small for $c_R \sim -1/2$. From Eq.~(\ref{TtopKK}),
for $c_R \approx 0$,   
$0.3 \stackrel{<}{\sim} c_L \stackrel{<}{\sim} 0.4$
and gauge/$t_L$ KK masses $\sim 3-4$ TeV, we get 
\begin{eqnarray}
T_{ \hbox{ KK top } }
& \sim & 
0.04 -0.3
\label{Trange2}
\end{eqnarray}
From Eqs.~(\ref{Trange2}), (\ref{Treq1}) and (\ref{Treq2}), 
we see that with radiatively generated $T$, 
the RS model agrees with the electroweak data for a sizable portion of the allowed
range of $T$ for light Higgs, whereas fit to data with heavy Higgs is possible
only at the upper limit of the allowed range of $T$.

We have not considered $c_L > 0.4$, since then the theory is no longer weakly
coupled at the KK mass scale,
the strong coupling scale (from loop corrections
to Higgs couplings due to top loop) estimated as: 
\begin{eqnarray}
\Lambda_t & \sim & z^{ -1 }_v \frac{ 4 \pi }{ \lambda_{ t \; 5 } k }  \nonumber \\
 & \sim & 4 m_{ t^{ (1) } } \sqrt{ ( 1 - 2 c_L ) ( 1 - 2 c_R ) } 
\end{eqnarray}
Similarly, to keep the theory weakly coupled we have chosen
$c_R \stackrel{<}{\sim} 0$ (given that $c_L \stackrel{>}{\sim} 0.3$ to 
avoid excessive corrections to bottom couplings). 
See section \ref{disc} for further discussion.

\section{Collider signals}
\label{signal}
Our model has rather distinctive phenomenology as follows.
As discussed earlier, the Higgs couplings to 
electroweak gauge KK modes are enhanced (compared to that of {\em zero}-modes)
by $\sim \sqrt{k \pi r_c}$ as expected from their CFT dual interpretation as strongly
coupled composites (see section \ref{CFT}). 
Thus,
{\em longitudinal} $W,Z$ (eaten Higgs component) fusion into electroweak gauge KK modes
(with masses $\sim$ few TeV)
is enhanced. In turn,
these KK modes have sizable decay widths to {\em longitudinal} $W/Z$'s:  
\begin{eqnarray}
W_{long.} \; Z_{long.} \; \hbox{and} \;
W_{long.} \; W_{long.} & \stackrel{ g \sqrt{ k \pi r_c } }{ \longrightarrow } 
& W^{ \pm \; (n) }, Z^{ (n) }, \tilde{W}^{ \pm \; (n) }, Z^{ \prime \; (n) } \nonumber \\
 & \stackrel{ g \sqrt{k \pi r_c} }{ \longrightarrow } 
& W_{long.} \; Z_{long.} \; \hbox{and} \; W_{long.} \; W_{long.}
\end{eqnarray}
Note that considerably below the energies of these resonances in
longitudinal $W/Z$ scattering, the growth of the cross section is softened 
by Higgs exchange.

There are also unique signals involving fermion modes. For example,
$t_R$ is strongly coupled to gluon and {\em right}-handed $\tilde{W}$ KK modes
since its wavefunction is localized near TeV brane, leading to 
$k \pi r_c$-enhanced production
of gluon and $\tilde{W}$ KK modes  
through gluon fusion via $t_R$ {\em loop}. Conversely,
gluon and {\em right}-handed $\tilde{W}$ KK modes
have strong decays to $t_R$:
\begin{eqnarray}
\hbox{gluon} + \hbox{gluon} & \stackrel{ 
\; \hbox{ 
top loop } }{ \longrightarrow } & \hbox{gluon}^{ (n) }, Z^{ (n) }, Z^{ \prime \; (n) }  
\nonumber \\
 & \stackrel{ g \sqrt{ k \pi r_c } }{ \longrightarrow }  & \bar{t}_R t_R 
\end{eqnarray} 
Again, this is expected since
in the dual CFT interpretation (see section \ref{CFT})
$t_R$ has a large admixture of CFT composites. 

Another interesting signal arises from enhanced
Higgs-$t_R^{ (0) }$-$b^{ (n) }_L$ coupling $\sim \lambda_{ t }
f (c_L) \sim \sqrt{10}$ (for $c_L \sim 0.4$),
leading to $b_L^{ (n) }$ production
by 
{\em longitudinal} $W$-$t_R^{ (0) }$ fusion:
\begin{eqnarray}
t_R \; W_{long.} & \stackrel{ f (c_L) \lambda_t }{\rightarrow}  & b_L^{ (n) }.
\end{eqnarray}
%

\section{CFT interpretation}
\label{CFT}

Our model 
is dual \cite{rscft}, in the sense of the AdS/CFT correspondence \cite{adscft}, to 
a strongly coupled large-$N$
$4D$ CFT 
with $SU(3)_c \times SU(2)_L \times SU(2)_R \times U(1)_{ B - L } $
global symmetry whose $SU(3)_c \times SU(2)_L \times U(1)_Y$ subgroup is gauged.
Higgs on TeV brane 
corresponds to a composite of the CFT responsible for spontaneous
breaking of $SU(2)_L \times SU(2)_R$ symmetry.
That is, our model is dual to a particular type of
a
composite Higgs model. 
In the dual interpretation, the hierarchy problem is solved by
this compositeness as opposed to any symmetry.

In the dual picture,
the $S$ parameter 
arises from exchange of
spin-$1$ CFT composites
(``techni-$\rho$'s'')
for which one would naively expect $S \sim
16 \pi v^2 / m_{ \rho }^2$ \cite{pt} 
-- this agrees
roughly with Eq.~(\ref{SRS}). 
However, because we have a Higgs in our model, $m_{ \rho }$ is {\em not}
tied to the weak scale and can be made heavy enough to adequately suppress the $S$ parameter.

The dual interpretation of light fermions
with $c > 1/2$ is as follows. 
We have fundamental fermions external to the CFT,
coupled to {\em ir}relevant fermionic CFT operators so that the mixing of the external
fermion with the CFT composites is small, i.e., the resulting
massless state which is the SM fermion is mostly fundamental.
Yukawa couplings to composite Higgs must go through this small mixing
and are therefore also small.
Also, because
SM fermions are mostly fundamental with small mixing to CFT,
higher-dimensional fermion-{\em Higgs} operators
as in Eq.~(\ref{fermionoperator}) are highly suppressed for light fermions.
The small mixing with CFT sector also naturally suppresses unwanted FCNC's.

For the third generation fermions, one can therefore
see a tension as follows. To obtain $\lambda _{t} \sim 1$,
it is clear that the top should couple to a relevant CFT operator
(dual to $c < 1/2$), i.e, fundamental
top quark should mix
appreciably with CFT composites in order for the 
SM top
to have $O(1)$ coupling to the composite Higgs.
However, if CFT operator coupled to fundamental
{\em left}-handed top and 
hence $b_L$ is {\em relevant}, then
the fundamental $b_L$ mixes 
substantially with the CFT composites
and induces
higher-dimensional bottom-Higgs operator in Eq.~(\ref{fermionoperator})
contributing to $Z \rightarrow \bar{b}_L b_L$.
To be consistent with $Z \rightarrow \bar{b}_L b_L$ data,
$b_L$ coupling to CFT operator can be at most {\em mildly} relevant.
But, then, in order to get the top Yukawa coupling, fundamental
$t_R$ has to couple to a more relevant operator. That is, 
the SM 
$t_R$ must contain sizable admixture of CFT composites.
This mechanism
for generating Yukawa hierarchies is similar to the proposal of reference
\cite{kaplan}, but translated into a CFT context here.

The central feature of the dual CFT that suppresses the $T$ parameter is exact
(in scenario II) custodial isospin from $SU(2)_R$ symmetry.
The dual interpretation of breaking of
$SU(2)_R$ by UV brane boundary condition is that 
this custodial isospin symmetry is {\em not} 
fully gauged. Rather, it is a symmetry of the CFT Higgs sector
when isolated from fundamental fermions and gauge fields.
These fundamental fermions and their couplings to the CFT explicitly break global $SU(2)_R$.


Of these custodial-isospin violating effects
due to fundamental fields, 
the largest is due to 
coupling of fundamental $t_R$ to {\em just} a isospin component of a moderately
{\em relevant} CFT operator.
As a result, the CFT 
flows to a {\em new} fixed point
which {\em violates} custodial-isospin at {\em sub}-leading
order in $N$.
This results in suppression of the $T$ parameter 
by
$1/N$ in scenario II, dual to generating
$T$ at {\em loop} order in the RS model.

The dual of scenario I with small bulk breaking of $SU(2)_R$
is that
the CFT Higgs sector has {\em approximate} custodial isospin (global) symmetry. 
Therefore, the $T$ parameter is suppressed,
but non-zero and can dominate over top-induced contribution discussed above.


\section{Discussion and Outlook}
\label{disc}
We have studied the precision electroweak fit in a RS1 scenario with
gauge fields and fermions in the bulk. In order to soften the $T$-parameter
constraints we have enhanced the electroweak gauge structure to
$SU(2)_L\times SU(2)_R\times U(1)_{B-L}$, recovering
the usual gauge group via Planck brane boundary conditions and Higgsing.
This model
has a natural exact
(approximate) custodial isospin {\it gauge} symmetry in the {\em bulk}
that renders
the 
$T$-parameter zero (small) at leading order in our weak/KK-scale expansion, 
compared with the excessive values obtained in earlier studies without 
bulk custodial isospin. 
Localizing the light
fermions near the Planck brane decreases the $S$-parameter
so the electroweak fit is possible for KK masses of a few TeV.
This type of bulk localization also allows us to incorporate the attractive 
mechanism for generating flavor structure of Refs. \cite{gp} \cite{huber}.

We have then complete and 
realistic models without any large hierarchies among input parameters. 
The collider
signals of our model are also quite distinctive and arise from the enhanced 
couplings of the top quark and longitudinal $W$'s and $Z$'s to KK modes.

Important insight into our model comes in the light of the 
AdS/CFT correspondence.  Our model is
dual to a strongly coupled CFT Higgs sector enjoying a custodial {\em global} 
symmetry, of which the minimal Higgs is a
composite arising after conformal invariance is broken,  
with couplings to fundamental SM gauge and fermion fields which 
necessarily violate custodial isospin. The large top mass is correlated 
with the right-handed top having a large admixture of a composite fermion 
within it, affecting its couplings.

The strong coupling 
scale (especially in our Scenario II) in the top-Higgs sector
is not much larger than the
scale of the first KK masses. The interpretation of this in effective field 
theory is that above this scale the Higgs-top physics is 
sensitive to the detailed structure of the IR brane, which here we treated 
as point-like in the extra dimension.
However this does not affect
the key results of our paper because bulk couplings have a higher 
strong-coupling scale.
Nevertheless it would be interesting in the future
to introduce explicit brane size and 
structure (and therefore structure in the Higgs
system)
within RS effective field theory so as to raise the scale of strong coupling 
on the IR brane to that of the bulk. This appears quite feasible.

Another aspect of
our scenario is the remnant fine tuning
needed in order to get a light Higgs compared to KK masses. It is
important to note that there are two contributions to the Higgs mass
at loop level, one coming from the zero modes and another from KK
modes. The latter contribution dominates 
since the KK couplings are enhanced over the zero mode couplings. 
We estimate fine tuning to be
of the order of $1\%$ in mass-squared from the top-Higgs couplings. 
We suspect the tuning can be made much milder by introducing symmetry 
protection for the Higgs from KK modes (as opposed to zero-modes which is 
inescapable). In particular the ideas of references \cite{cnp} and 
\cite{partlysusy} seem promising and attractive.

Finally one can ask whether our models can be embedded into a GUT
theory along the lines proposed in reference \cite{us}. Since we
have enlarged the gauge structure, minimal $SU(5)$ will not work as in 
reference 
\cite{us}, but
both  $SU(2)_R$ and $SU(5)$ can be easily embedded into $SO(10)$, 
and a realistic model incorporating both symmetries seems feasible.

\section*{Acknowledgments}

The work of
K.~A.~was supported by the Leon Madansky fellowship
and NSF Grant P420D3620414350.
The work of A.~D.~was supported by NSF 
Grants P420D3620414350 
and
P420D3620434350. 
The work of M.~M.~was supported in part by NSF Grant P420D3620434350.
The work of 
R.~S.~was supported by 
NSF Grant P420D3620434350. 
We thank Marcela Carena, Csaba Csaki, Tony Gherghetta, 
Christophe Grojean,
Ian Hinchliffe,
David E.~Kaplan,
Markus Luty, Konstantin Matchev, Gilad Perez, Michael Peskin, Alex Pomarol, 
Eduardo Ponton,  Mariano Quiros, Lisa Randall,
Riccardo Rattazzi, Martin Schmaltz, Tim Tait, John Terning
and Carlos Wagner for discussions and 
the Aspen Center
for Physics for hospitality during the completion of this work.

\section*{Appendix}
\begin{appendix}


\section{5D Gauge Propagators}
\label{5Dprop}
We use 
$5D$
mixed position-momentum 
space propagators in Feynman gauge. 
The general propagator from $z$ to $z^{\prime}$ for $(+,+)$ boundary conditions with momentum $p$ 
is (up to tensor structure) \cite{lisa}:
\begin{eqnarray}
G^{ 5 D }_{ p (++) } (u, v) & = & \frac{\pi}{2} 
\frac{k u v}{ \big[ - Y_0 (p z_h) J_0 (p z_v) + J_0 (p z_h)
Y_0 (p z_v) \big] } \nonumber \\
 & & \times 
\big[ - Y_0 (p z_h) J_1 ( pu ) + J_0 (p z_h) Y_1 
(pu) \big] \big[ -Y_0 (p z_v) J_1 (pv) + J_0 (p z_v) Y_1 (pv) \big], \nonumber \\ 
\label{pp}
\end{eqnarray} 
where
$u = min(z, z^{\prime})$ and $v = max(z, z^{\prime})$.


For $(-,+)$ 
boundary condition with bulk mass $M$, the propagator is
\begin{eqnarray}
G^{ 5 D }_{ p (-+) } ( u, v, M )  & = & \frac{\pi}{2} 
\frac{k u v}{ \big[ - Y_{ \nu } (p z_h) \tilde{ J }_{ \nu } 
(p z_v) - J_{ \nu } (p z_h)
\tilde{ Y}_{ \nu } (p z_v) \big] } \nonumber \\
 & & \times 
\big[ - Y_{ \nu } (p z_h) J_{ \nu } ( pu ) + J_{ \nu } (p z_h) Y_{ \nu} 
(pu) \big] \big[ -\tilde{Y}
_{ \nu } (p z_v) J_{ \nu } (pv) + \tilde{J}_{ \nu } (p z_v) Y_{ \nu } 
(pv) \big], \nonumber \\  
\label{mp}
\end{eqnarray}
where $\nu = \sqrt{ 1 + M^2 / k^2 }$ and $\tilde{J}_{ \nu } (x) \equiv ( 1 - \nu )/ x \;
J_{ \nu }
(x)
+ J_{ \nu - 1 } (x)$.


\section{KK Sum to 5D Propagator Conversion}
\label{KKsum5Dprop}

We expand corrections to the zero mode gauge propagators 
in $v z_v$ (see section \ref{STgauge}), i.e.,  
we work in the                 
insertion approximation for the Higgs vev (see discussion at the end of section 
\ref{formalism}).   
We isolate from $\mathcal{L}_{\mathrm{IR}}$ 
all terms containing only zero modes.  The zero modes reproduce electroweak symmetry 
breaking at order $v^2$. In particular, the zero modes yield $M_W^2=M_Z^2 \cos^2 \theta_W$, 
where $\theta_W$ is the weak mixing angle.  

As in the standard framework for oblique corrections 
\cite{pt}, 
we will consider diagrams with $W^{a \, (0)}$'s on the external lines.  
The Higgs vev mixes $W^{a \, (0)}$ with its own KK modes, 
and also the $\widetilde{W}^a$, $Z^{\prime}$ and $B$ mass eigenstates.  
The KK expansion for the dimension $\frac{3}{2}$ field 
is 
\be
W_\mu^{a \, 5D}(x,y)= \sum_n W^{a \,(n)}_\mu(x) \frac{f_n(y)}{\sqrt{ \pi r_c}}. 
\ee
Using this expansion and inserting $\langle H \rangle$, we find
\be
\label{4dkkandvev}
\mathcal{L}_{\mathrm{IR}} \supset g_{5} g_{5i} \frac{v^4}{16}   \sum_{n,m} W^{a (n)}_{\mu} 
X^{i \, (m) \, \mu} \frac{f^a_n(z_v)}{\sqrt{ \pi r_c}} \frac{f^i_m(z_v)}{\sqrt{ \pi r_c}}.
\ee
If $a=3$,  $X^{i \, (m) \, \mu}$ is 
$W^{3 \, (m) \, \mu}$,  $Z^{\prime \, (m) \, \mu}$, or 
$B^{(m) \, \mu}$.  If $a=1$ or $2$, $X^{i \, (m) \, \mu}$ is 
$W^{1,2 \, (m) \, \mu}$ or $\widetilde{W}^{1,2 \, (m) \, \mu}$.  
$g_{5i}$ is the appropriate coupling in each case, and $f^i_n(z_v)$ 
is evaluated at the IR brane.  

With a left handed zero mode on each external line, $f^a_0(z_v)=1$
since zero-mode has a flat profile.   
Using (\ref{4dkkandvev}) and Feynman gauge, diagrams like Fig.~\ref{Ttree} are
\be
\label{5dterms}
  g^2_{5} g^2_{5i} \frac{v^4}{16} \sum_{n}  \frac{i}{ (\pi r_c)^2} \frac{f^i_n(z_v) f^i_n(z_v)}
{q^2-m_n^2} \eta^{\mu \nu}.
\ee
Here, $m_n$ is the mass of the $n^{\mathrm{th}}$ KK mode, and $q^2$ is the four momentum in the KK propagator. 

Summing over eigenfunctions,  Fig.~\ref{Ttree} is then
\be
\label{kktoG}
g^2 g^2_{5i} \frac{v^4}{16} \sum_{n}  \frac{i}{ \pi r_c} \frac{f^i_n(z_v)f^i_n(z_v)} 
{q^2-m_n^2}=i \frac{v^4}{16} g^2 g^2_{5i} G^{5D}_{q \, i}(z_v, z_v).
\ee
$G^{5D}_{q \, i}(z_v, z_v)$ is 
given in equations (\ref{pp}) and (\ref{mp}))
and $g = g_5 / \sqrt{ \pi r_c }$ is the $4D$/zero-mode gauge coupling. 
Choosing to write the graphs 
in terms of $G^{5D}_{q \, i}(z_v, z_v)$ automatically does the sum over KK modes and considers 
the different boundary conditions on the $f^i_n(z_v)$ for each field.  With this prescription, 
the Feynman rules for Fig.~ \ref{Ttree} are simply $g g_{5i} \frac{v^2}{4} $ at each vertex, and 
$i G^{5D}_{q\, i}(z_v, z_v)$ for the gauge boson propagator.

The propagator for the zero-mode is:
\begin{equation}
G^{ (0) }_ { q (++) } = \frac{1}{ q^2 \pi r_c }
\end{equation}
There is no zero-mode for $(-,+)$ boundary condition.

For the tree-level contribution to $T$ from gauge-Higgs sector,
we need 
propagator at zero momentum 
with {\em zero}-mode {\em subtracted}. For $(+,+)$ boundary conditions, this is
(up to $O \left( z_h^2 / z_v^2 \right)$)
\begin{eqnarray}
G^{ 5D }_{q = 0 
(+,+)} (z_v, z_v) - G^{ (0) }_ { q (++) } 
& = & \frac{1}{ \pi r_c } \left( \frac{ z_v^2 }{2}
- \frac{ z_v^2 k \pi r_c }{2} - \frac{ z_v^2 }{ 4 k \pi r_c } \right)
\label{IRIR}
\end{eqnarray}
and for $(-,+)$ boundary condition with a bulk mass $M$ (up to
$O \left( z_h^2 / z_v^2, \; M^4 / k^4 \right)$):
\begin{eqnarray}
G^{ 5D }_{q =0 
(-,+) } (z_v, z_v) & = & - \frac{ z_v^2 }{ 2
z_h } \left( 1 - \frac{ M^2 }{ 4 k^2 } \right)
\end{eqnarray}
%


\section{Fermionic operators}
\label{appfermion}

As in the calculation of 
the diagram in Fig.~\ref{Ttree}, we convert the sum over KK modes
in Figs.~\ref{compositeness} and \ref{Stree} into a $5D$ propagator.
For compositeness Feynman diagram (Fig.~\ref{compositeness}), we need propagator at zero
momentum 
with {\em zero}-mode {\em subtracted}. For 
$(+,+)$ boundary conditions this is
\begin{eqnarray}
G^{ 5D }_{q =0 
(++) } (u, v) - G^{ (0) }_{ q (++) } 
& \approx & \frac{1}{\pi r_c} \left(
\frac{ u^2 }{4}  - \frac{ u^2 \log \frac{u}{ z_h } }{2} + \frac{ v^2 }{4}    
- \frac{ v^2 \log \frac{v}{ z_v } }{2} 
- \frac{z_v^2}{4 k \pi r_c} 
\right)
\nonumber \\
\end{eqnarray}

Similarly, 
for fermion-Higgs operator Feynman diagram (Fig.~\ref{Stree}), we
need propagator from TeV brane to $z$
for zero momentum 
(again, with zero-mode subtracted):
\begin{eqnarray}
G^{ 5D }_{q =0 (++) 
} (z, z_v) - G^{ (0) }_{ q (++) } & \approx & \frac{1}{\pi r_c} \left(
\frac{ z^2 }{4} - \frac{ z^2 \log k z }{2} + \frac{ z_v^2 }{4}
- \frac{z_v^2}{4 k \pi r_c} 
\right)
\nonumber \\
\label{ztoTeV++}
\end{eqnarray}
for $(+,+)$ boundary condition
and
\begin{eqnarray}
G^{ 5D }_{ q =0
} (z, z_v) & \approx &
\frac{ z_h }{2} \left( 1 - \frac{ z^2 }{ z_h^2 } \right)
\end{eqnarray}
for $(-, +)$ boundary condition with $M = 0$.

The $c$-dependent
wavefunction of fermion zero-mode is (see, for example, \cite{gp, song}):
\begin{eqnarray}
\Psi ( x, z ) & \ni & \psi ^{ (0) } (x) \chi_0 ( c, z ), \; 
\hbox{where} \nonumber \\
\chi_0 (c, z) & = & \sqrt{ 
\frac{1-2c}{ z_h \left( e^{ k \pi r_c ( 1-2c) }
-1 \right) } } \left( \frac{z}{z_h} \right)^{2-c}
\label{zeromode}
\end{eqnarray}
%


\section{Gauge KK masses and couplings}
\label{KKTeVcoupling}

The
masses of gauge KK modes with
$(+,+)$ boundary condition
are given by 
\begin{eqnarray}
\frac{ J_0 \left( m_{ gauge }^{ (n) } z_h \right) }{ Y_0 \left( m_{ gauge }
^{ (n) } z_h \right) } & = & 
\frac{  J_0 \left( m_{ gauge }^{ (n) } z_v \right) }{  Y_0 \left( m_{ gauge }^{ (n) } z_v
\right) } 
\end{eqnarray}
so that, for $m^{ (n) }_{ gauge } z_h \ll 1$, we get
$m^{ (n) }_{ gauge } z_v \approx $ zeroes of $J_0 + O \left( 1 / 
\big[ 
\log m^{ (n) }_{ gauge } z_h 
\big] \right)$. In particular,
the mass of the lightest gauge KK mode is given by
\begin{equation}
m^{ (1) }_{ gauge } \approx 2.45 z_v
\label{lightestgaugeKK}
\end{equation}

The
masses of gauge KK modes with
$(-,+)$ boundary condition and a bulk mass $M$
are given by 
\begin{eqnarray}
\frac{ J_{ \nu } \left( m_{ gauge }^{ (n) } z_h \right) }{ Y_{ \nu } 
\left( m_{ gauge }
^{ (n) } z_h \right) } & = & 
\frac{  \tilde{J}_{ \nu } \left( m_{ gauge }^{ (n) } z_v \right) }{  
\tilde{Y}_{ \nu } \left( m_{ gauge }^{ (n) } z_v
\right) } 
\end{eqnarray}
so that, for $m^{ (n) }_{ gauge } z_h \ll 1$, we get
$m^{ (n) }_{ gauge } z_v \approx $ zeroes of $J_0 + 
M^2 / \left( 4 k^2 \right) + O \left( \big[ z_h m^{ (n) }_{ gauge }
\big]^2, \; M^4 / k^4
\right)$.

We compute 
couplings of gauge KK modes to Higgs/TeV-brane localized fields
by writing
the TeV-brane-to-TeV-brane propagator with
zero-mode subtracted (at zero-momentum)
as a sum over KK modes
multiplied by respective couplings (as in Eq.~(\ref{kktoG})):
\begin{eqnarray}
\sum_n \frac{ \left( f^i_n \right) ^2 ( z_v ) }{ \pi r_c } 
\frac{1}{ m_{gauge}^{ (n) \; 2} } \approx \frac{ z_v^2 }{ 2 z_h }
\end{eqnarray}
using Eq.~(\ref{IRIR}). Using the above spectrum for KK modes gives
$f^i_n (z_v) \approx \sqrt{ 2 k \pi r_c } f_0 (z_v) $
(for both $(+,+)$ and $(-,+)$ boundary conditions), i.e.,
coupling of gauge KK modes to TeV-brane localized fields
is enhanced compared to that of zero-mode
by $\sqrt{ 2 k \pi r_c }$.


\section{Spectrum and couplings to Higgs (at TeV brane)
of $t_R$ and $(t, b)_L$ $(+,+)$}
\label{spectrum++}

There is a chiral zero-mode and
the masses of KK modes, $m_n$ are given by (see, for example, \cite{choi}):
\begin{eqnarray}
\frac{ J_{ \alpha \mp 1} \left( m_n z_h \right) }
{ Y_{ \alpha \mp 1} \left( m_n z_h \right) } & = & 
\frac{ J_{ \alpha \mp 1} \left( m_n z_v \right)  }
{ Y_{ \alpha \mp 1} \left( m_n z_v \right) } \equiv - b_{ \alpha } (m_n), 
\label{tR++}
\end{eqnarray}
where the upper (lower) signs are for $c > -1/2$ ($c < -1/2$) and
$\alpha = | c + 1/2 |$.
We will need masses of lightest KK modes only so that henceforth we assume
$m_n z_h \ll 1$.
 
For $-1/2 < c < 1/2$, we get $-1 < ( \alpha - 1 ) < 0$ 
so that, using 
$Y_{ \nu } = 1 / \sin \nu \pi \; \left( 
J_{ \nu } \cos \nu \pi - J_{ - \nu } \right)$, we see that
$Y_ { \alpha - 1 } \left( m_n z_h \right) 
\propto \left( m_n z_h \right)^{ \alpha - 1 }$  
just like $J_{ \alpha - 1 } \left( m_n z_h \right)$.
Thus, LHS of Eq.~(\ref{tR++}) does not have a ``convenient'' limit.
So, we use the above relation for $Y_{ \nu }$
to 
show that 
$Y_{ \alpha - 1 }$ 
can be 
replaced by $J_{ - \alpha + 1 }$ (on both sides)
in the above 
equation\footnote{Or we can use
$J_{ - \alpha }$ instead of $Y_{ \alpha }$ as the Bessel function
of the second kind in the wavefunctions of KK modes
to get the same result.}.
Then, 
for $-1/2 < c < 1/2 - \epsilon$ (for a suitable $\epsilon$: see below), 
LHS of Eq.~(\ref{tR++}) $\sim  \left( m_n z_h \right) ^{ 2 ( c-1/2 ) } > 
1/ \left( m_n z_h \right)^{ 2 \epsilon} \gg 1$ 
so that
$m_n z_v \approx$ zeroes of $J_{ 1 - \alpha } = J_{ -c + 1/2 }$. Therefore, the
lightest KK mass
$\sim z_v^{-1}$ so that, in order to validate the assumption that LHS of Eq.~(\ref{tR++}) $ \gg
1$,
we need $\left( z_h / z_v \right)^{\epsilon} \ll 1$: we see that $\epsilon 
\stackrel{>}{\sim} 0.1$ suffices. Whereas, 
for $c < -1/2 - \epsilon$, 
LHS of Eq.~(\ref{tR++}) $\sim \left( m_n z_h \right)^{ 2 ( \alpha + 1 ) } \sim 0$ 
and so
we get 
$m_n z_v \approx$ zeroes of $J_{ \alpha + 1 } = J_{ -c +1/2 }$.

The wavefunctions of KK modes are given by 
(see,
for example, \cite{gp, song}):
\begin{eqnarray}
\Psi (x, z) & \ni & \frac{ 
\left( \frac{z}{ z_h } \right)^2
}{ \sqrt{\pi r_c} }
\sum_{ n \neq 0 } \psi^{(n)} (x) \chi_n 
(z), 
\end{eqnarray}
\begin{eqnarray}
\chi_n 
(z)
& = & \frac{ 
\sqrt{ \frac{z}{ z_h } }
}{ N_n } 
\big[ J_{ \alpha } \left( m_n 
z
\right) + 
b_{ \alpha } (m_n) 
Y_{ \alpha } \left( m_n 
z
\right)
\big],
\end{eqnarray}
\begin{eqnarray}
\frac{1}{ \pi r_c } \int_{
z_h
}^{ 
z_v
} d z
\chi^2_n 
(z)
& = & 1,
\end{eqnarray}
and 
\begin{eqnarray}
N_n^2 & = & \frac{1}{ 2 \pi r_c z_h } 
\Big[ z_v^2 \big[ J_{ \alpha } \left( m_n z_v \right) 
+ b_{ \alpha } (m_n) Y_{ \alpha } \left( m_n z_v \right) \big]^2 - 
z_h^2 \big[ J_{ \alpha } \left( m_n z_h \right) 
+ b_{ \alpha } (m_n) Y_{ \alpha } \left( m_n z_h \right) \big]^2 \Big].
\nonumber \\
\end{eqnarray}

In particular, we need coupling of KK modes to Higgs to determine mass
insertions in Fig.~\ref{Ttoploop2}, i.e.,
wavefunction of KK modes {\em at TeV} brane:
\begin{eqnarray}
\Psi (x , z_v ) & = & \sum_n \frac{ \psi^{(n)} (x) }{ \sqrt{ 1 - \delta_n } } 
\frac{ \sqrt{2} z_v^{ \frac{3}{2} } }{ z_h^2 },
\label{KKatTeV}
\end{eqnarray}
where
\begin{eqnarray}
\delta_n & = & \frac{ z_h^2 
\big[ J_{ \alpha } \left( m_n z_h \right) + 
b_{ \alpha } (m_n) Y_{ \alpha } \left( m_n z_h \right) 
\big]^2 }
{ z_v^2 \big[ J_{ \alpha } \left( m_n z_v \right) + 
b_{ \alpha } (m_n) Y_{ \alpha } \left( m_n z_v \right) 
\big]^2 } 
\end{eqnarray}
We get $\delta_n \approx 0$ for 
$m_n z_h \ll 1$.

It is useful to define $f (c)$ 
to be ratio of wavefunction {\em at TeV brane} of KK mode
and zero-mode (using Eqns. (\ref{zeromode}) and (\ref{KKatTeV})):
\begin{eqnarray}
f (c) & \approx & \sqrt{ \frac{2}{ 
1 - 2 c 
} }
\label{f}
\end{eqnarray}
for $c < 1/2 - \epsilon$ and using $\delta _n \approx 0$.
Thus,
$t_L^{ ( n \neq 0 ) } t_R^{ (0) }$ ``mass insertion'' 
is $m_t f (c_L)$ and 
$t_L^{( n \neq 0 )} 
t_R^{( m \neq 0 )}$ mass insertion is $m_t f (c_L) f (c_R)$ -- these mass
insertions are used in the calculation in section \ref{Ttop}.


\section{Spectrum and 
couplings to Higgs
(at TeV brane) of $b^{\prime}_R$ $(-,+)$}
\label{spectrum-+}

For $(-,+)$ boundary condition,
there is no zero-mode
and the spectrum of KK modes is given by \cite{choi}:
\begin{eqnarray}
\frac{ J_{ \alpha } \left( m_n z_h \right) }
{ Y_{ \alpha } \left( m_n z_h \right) } & = & 
\frac{ J_{ \alpha \mp 1} \left( m_n z_v \right) }
{ Y_{ \alpha \mp 1} \left( m_n z_v \right) } \equiv - b_{ \alpha } (m_n). 
\label{bR-+}
\end{eqnarray}

An analysis similar to the one above shows that, for $c > -1/2 + \epsilon$, 
$m_n z_v \approx$ zeroes of $J_{ c-1/2 }$,
whereas,
for $c < -1/2 - \epsilon$, 
$m_n z_v \approx$ zeroes of $J_{ -c +1/2 }$.
 
In addition, for $c < -1/2 - \epsilon$,
we can show that there is a 
mode much lighter than $1/z_v$
(when arguments of both LHS and RHS of Eq.~(\ref{bR-+}) are small) given by
$m_n z_v \approx 2 \sqrt{ \alpha ( \alpha + 1 ) }  
\left( z_h/z_v \right)^{ \alpha } \ll z_v^{ -1 }$.

The expressions for wavefunctions of KK modes are similar to those 
for $(+,+)$ KK modes, except:
\begin{eqnarray}
N^2_n & = & 
\frac{1}{ 2 \pi r_c z_h } 
\Big[ z_v^2 \big[ J_{ \alpha } \left( m_n z_v \right) 
+ b_{ \alpha } (m_n) Y_{ \alpha } \left( m_n z_v \right) \big]^2 - 
z_h^2 \big[ J_{ \alpha \mp 1 } \left( m_n z_h \right) 
+ b_{ \alpha } (m_n) Y_{ \alpha \mp 1 } \left( m_n z_h \right) \big]^2 \Big]
\nonumber \\\end{eqnarray}
so that
\begin{eqnarray}
\delta^{ \prime }_n & = & \frac{ z_h^2 
\big[ J_{ \alpha \mp 1 } \left( m_n z_h \right) + 
b_{ \alpha } (m_n) Y_{ \alpha \mp 1 } \left( m_n z_h \right) 
\big]^2 }
{ z_v^2 \big[ J_{ \alpha } \left( m_n z_v \right) + 
b_{ \alpha } (m_n) Y_{ \alpha } \left( m_n z_v \right) 
\big]^2 }. 
\end{eqnarray}

For $c > -1/2 + \epsilon$, we get
$\delta^{ \prime }_n \approx 0$ so that wavefunction {\em at TeV} brane is {\em same}
as for $(+,+)$ KK modes. Thus, $b_L^{ ( m \neq 0 ) } b^{ \prime \; (n) }_R$ 
mass insertion is $m_t 
f (c_L) f ( c_R )$ -- these mass
insertions are used in the calculation in section \ref{Ttop}.

For $c < -1/2 - \epsilon$, 
for the ``very light'' mode, we get 
$\delta^{ \prime } \approx (c+1/2)/(c-1/2)$ so that
(using Eqs.~(\ref{zeromode}) and (\ref{KKatTeV})) 
very light mode has the  
{\em same}
wavefunction on TeV brane as {\em zero}-mode of $t_R$, whereas 
$\delta^{ \prime } _n \approx 0$ for the other modes of
$b^{ \prime }_R$. 

\end{appendix}

\end{document}